\documentclass{statsoc}
\pdfoutput=1

\usepackage{graphicx}
\usepackage{amsmath,amsfonts,amssymb}
\usepackage{rotating}
\usepackage{amsfonts}
\usepackage{subcaption}

\usepackage[rgb]{xcolor}
\definecolor{comment}{rgb}{255, 0, 0}

\usepackage{proofread}

\usepackage[ruled,vlined]{algorithm2e}

\usepackage[a4paper]{geometry}

\newcommand{\beginsupplement}{%
        \setcounter{table}{0}
        \renewcommand{\thetable}{S\arabic{table}}%
        \setcounter{figure}{0}
        \renewcommand{\thefigure}{S\arabic{figure}}%
        \setcounter{section}{0}
        \renewcommand{\thesection}{S\arabic{section}}%
     }

\title[Flexible noise processes]{Using flexible noise models to avoid noise model misspecification in inference of differential equation time series models}

\author{Richard Creswell}
\address{Department of Computer Science, University of Oxford, Oxford, United Kingdom}
\email{richard.creswell@hertford.ox.ac.uk}
\author{Ben Lambert}
\address{MRC Centre for Global Infectious Disease Analysis, School of Public Health, Imperial College London, United Kingdom}
\author{Chon Lok Lei}
\address{Department of Computer Science, University of Oxford, Oxford, United Kingdom}
\author{Martin Robinson}
\address{Department of Computer Science, University of Oxford, Oxford, United Kingdom}
\author[R. Creswell, B. Lambert, C. L. Lei, M. Robinson, and D. Gavaghan]{David Gavaghan}
\address{Department of Computer Science, University of Oxford, Oxford, United Kingdom}

\date{2020 March 02}

\begin{document}

\begin{abstract}
When modelling time series, it is common to decompose observed variation into a ``signal'' process, the process of interest, and ``noise'', representing nuisance factors that obfuscate the signal. To separate signal from noise, assumptions must be made about both parts of the system. If the signal process is incorrectly specified, our predictions using this model may generalise poorly; similarly, if the noise process is incorrectly specified, we can attribute too much or too little observed variation to the signal. With little justification, independent Gaussian noise is typically chosen, which defines a statistical model that is simple to implement but often misstates system uncertainty and may underestimate error autocorrelation. There are a range of alternative noise processes available but, in practice, none of these may be entirely appropriate, as actual noise may be better characterised as a time-varying mixture of these various types. Here, we consider systems where the signal is modelled with ordinary differential equations and present classes of flexible noise processes that adapt to a system's characteristics. Our noise models include a multivariate normal kernel where Gaussian processes allow for non-stationary persistence and variance, and nonparametric Bayesian models that partition time series into distinct blocks of separate noise structures. Across the scenarios we consider, these noise processes faithfully reproduce true system uncertainty: that is, parameter estimate uncertainty when doing inference using the correct noise model.  The models themselves and the methods for fitting them are scalable to large datasets and could help to ensure more appropriate quantification of uncertainty in a host of time series models.
\end{abstract}

\keywords{time series, ordinary differential equations, noise model, non-stationary processes, Bayesian inference, Gaussian process, Bayesian nonparametrics}

\section{Introduction}
Time series data are ubiquitous in many fields of scientific inquiry. In this paper, we focus on time series data that are assumed to obey a (potentially nonlinear) parametric model $f(t; \theta)$, a function of time $t$ and model parameters $\theta$. We model a noise-free trajectory $\{\bar y_i\}_{i=1}^N$ at time points $\{t_i\}_{i=1}^N$ according to,
\begin{equation}
\bar y_i = f(t_i; \theta).
\end{equation}
Often, scientific knowledge about a given system does not specify $f$ directly but rather suggests an ordinary differential equation (ODE) or system of ODEs with $f$ as their solution. In the following, we assume that these equations are numerically solvable, so that the mean or noise-free trajectory\footnote{In general, when $f$ is obtained by numerically solving ODEs, some numerical error may occur.} $f(t; \theta)$ is readily available for given values of $t$ and $\theta$: this is known as solving the ``forward problem''. Here, we consider the so-called ``inverse problem'': where, given a time series of noisy observations, the aim is to infer the values of $\theta$ that have generated the observations. A wide variety of inference tasks fall into this category, including parameter estimation for enzyme kinetics, biochemical pathways, and regulatory dynamics~\citep{milstein1981inverse, moles2003parameter, silk2011designing}. The Bayesian approach to the inverse problem, which we adopt in this paper, yields a posterior probability distribution over model parameters that conveys uncertainty in parameter estimates implied by data~\citep{gelman2013bayesian}.

When solving the forward problem, assumptions made about the form of the noise can substantially change estimated posterior uncertainty of $\theta$~\citep{autoregressivepaper}. Notably, when the noise model is misspecified, posterior variance in model parameters may be drastically underestimated or overestimated. Misspecification may also lead to biased estimates. The standard assumption of independent and identically distributed (IID) Gaussian noise is applicable in some cases, but there are many other possible forms. For example, consecutive observations may be correlated due to imperfections in measurement rather than the shape of the signal itself; the magnitude of measurement noise may scale with function values; there may be time periods with higher observation volatility due to environmental variation; or even a mixture of these various types of noise within a single time series. Non-Gaussian and non-IID noise is also likely to appear in cases of time series model misspecification: when the best available model does not coincide with the hypothetical true process which generated the data, regions of poor fit may be accompanied by residual autocorrelation and spikes in the magnitude of the noise terms.

In applied circumstances, the exact noise process is never known. Some form for the noise must therefore be assumed, with consequences for inference. Whatever choice is made should have some rational basis but be flexible enough to account for the particular sample of data to hand. In this vein, parametric models likely fall short and, instead, more adaptable non-parametric methods prosper. Here, we describe a number of non- or semi-parametric models for capturing noise processes that defy characterisation into existing boxes. Through a host of toy examples with predetermined noise processes, we show that parameter inference using our noise models faithfully reproduces the true posterior distributions; that is, those distributions that result when using the correct noise process. Our noise models, and the methods for fitting them, are designed to scale well with data size and could be used across a large range of ODE models for time series. To facilitate their use, we have designed our noise models and fitting routines to work with Pints software~\citep{clerx2019probabilistic} -- a general purpose Python library for fitting time series models, available from https://github.com/pints-team/pints. The code for this paper is available from https://github.com/rcw5890/flexnoise.

\begin{figure}[h!]
\centering
\includegraphics[width=\linewidth]{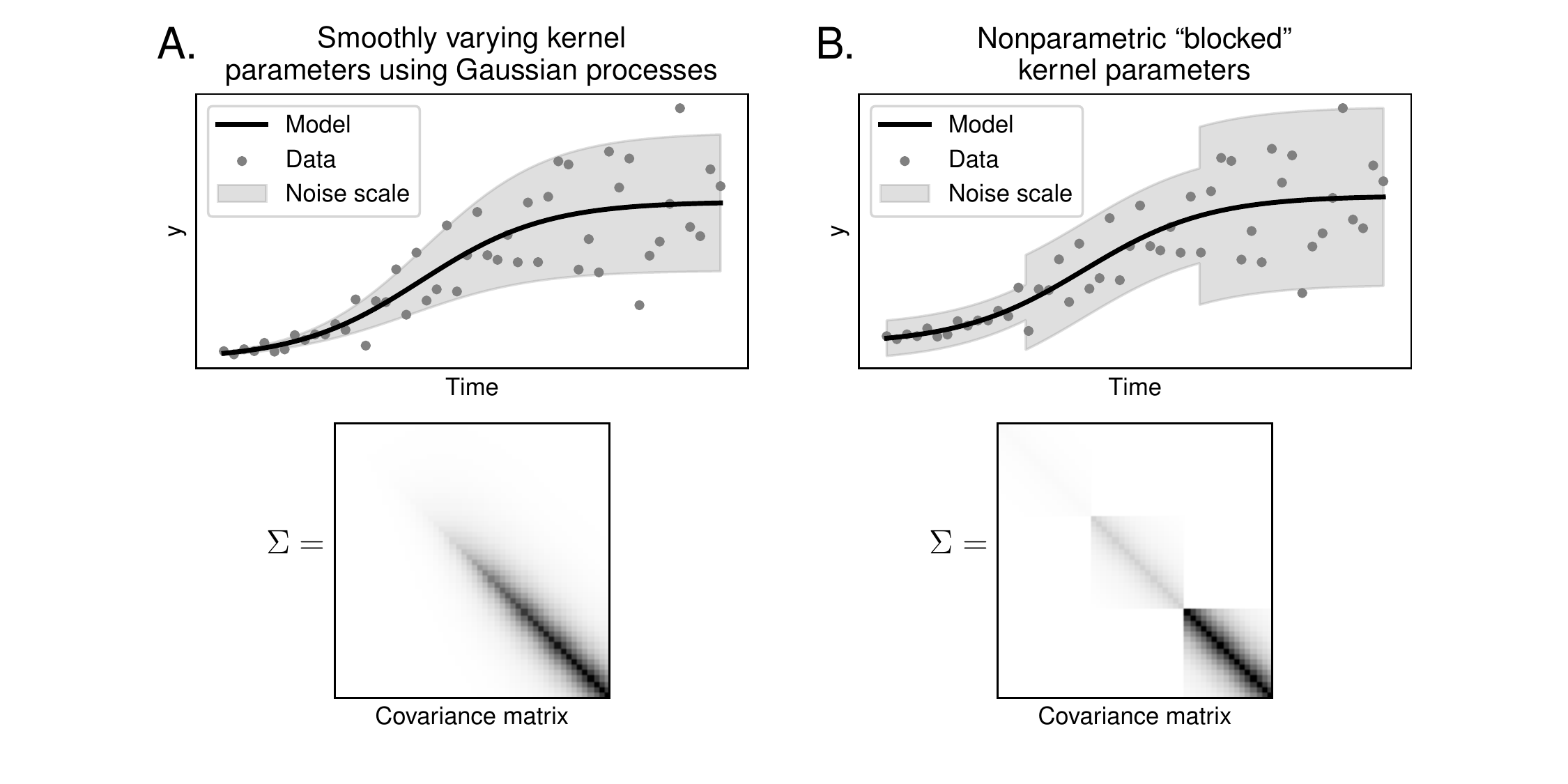}
\caption{\textbf{Two noise processes for time series modelling.} Panel (A) shows how non-stationary covariance kernels with continuously time-varying parameters can be used to learn the covariance matrix; and panel (B) shows how a covariance matrix can be built from non-overlapping constituent blocks.}
\label{fig:cartoon}
\end{figure}

Figure~\ref{fig:cartoon} gives an overview of the proposed noise model and the two different methods we use to generalise it to non-stationary noise. In both panels, time series data is illustrated with non-IID noise: the noise terms increase in magnitude as time proceeds. Panels A and B illustrate the two Bayesian methods we consider for learning appropriate covariance matrices. The first method, shown in panel A, uses a non-stationary covariance kernel whose parameters can vary continuously over time. As illustrated, this method is capable of learning a covariance matrix which steadily increases in magnitude along the diagonal. The next method, shown in panel B, divides time series into multiple regimes each with their own noise parameters. While both of these Bayesian methods allow a high degree of flexibility, they can be constrained via their prior distributions. Although we rely on a joint multivariate normal distribution for the data, and employ the Gaussian process in one of our noise processes, the methods presented here differ substantially from a Gaussian process regression, as we assume that the noise-free signal does not deviate from the given parametric model $f$. In~\S\ref{gp_comparison}, further discussion of the relationship between our methods and Gaussian processes is provided.

The remainder of this paper is organised as follows. In \S\ref{normal}, the multivariate normal distribution is introduced as a general model for noisy time series data, and we show how an appropriate covariance matrix can be learned from data using positive definite kernels. In \S\ref{gp} and \S\ref{blocks}, we describe two different methods to generalise these kernel methods to cases where the structure of noise changes over time. In \S\ref{gp}, we describe a method where the parameters of a kernel vary smoothly over time governed by Gaussian processes. In \S\ref{blocks}, we consider clustering methods which divide a time series into different regimes, with the kernel parameters taking different values in each of these. In \S\ref{performance}, we discuss performance considerations for long time series, and \S\ref{results} shows the application of our methods to real data from experiments conducted on the hERG potassium ion channel.

\section{The multivariate Gaussian likelihood for time series noise}  \label{normal}
The models we propose as flexible noise processes both depend on a suitably general distributional assumption governing the difference between observed and ODE-predicted data, and we use the multivariate normal. To learn the covariance matrices of the multivariate normal, we use positive definite kernel functions. This section introduces these concepts and shows how they can be used to correctly infer parameter posteriors for a time series model with stationary but non-IID noise.
\subsection{Description of multivariate likelihood}
The dataset consists of time points $\{t_i\}_{i=1}^N$ and corresponding noisy data $\{y_i\}_{i=1}^N$. A typical modelling assumption is to treat the noise on each data point as IID Gaussian with a variance parameter $\sigma^2$, so that,
\begin{equation}  \label{eq:1}
y_i = f(t_i; \theta) + \epsilon_i, \quad i=1,\dots, N,
\end{equation}
\begin{equation}  \label{eq:noise}
\epsilon_i \overset{\text{IID}}{\sim} \mathcal{N}(0, \sigma^2).
\end{equation}
Our first step is to generalize eq. \eqref{eq:noise} so that the variance of noise terms can vary (i.e. allow the noise to be non-identically distributed), and each noise realisation can be correlated with its neighbours (i.e. be non-independent). A multivariate Gaussian can handle both of these generalisations, where we model a random vector, $\mathbf{y} = (y_1, \dots, y_N)^\top$, as having a mean, $\mathbf{f}(\theta) = (f(t_1; \theta),\dots,f(t_N; \theta))^\top$,
\begin{equation}  \label{eq:2}
\mathbf{y} \sim \mathcal{N}(\mathbf{f}(\theta), \Sigma).
\end{equation}
For appropriate values of the covariance matrix $\Sigma$, this distributional assumption encompasses a wide variety of noise forms which may include correlated and heteroscedastic noise terms. For example, eq. \eqref{eq:2} could describe heteroscedastic noise which scales with the magnitude of the trajectory with $\Sigma=\text{diag}(\mathbf{f}(\theta) \sigma^2)$. For autocorrelated noise terms, $\Sigma$ would be dense.

\subsection{Learning the covariance matrix, $\Sigma$}
Multiple methods have been proposed for inference of covariance matrices~\citep{hoffbeck1996covariance, lam2009sparsistency, diggle1998nonparametric, bickel2008regularized, cai2011adaptive,schafer2005shrinkage}. A standard Bayesian approach places a prior on $\Sigma$ and infers it along with ODE model parameters, $\theta$. Typical choices for priors include the conjugate inverse-Wishart \citep{gelman2013bayesian, huang2013simple}, or a prior based around the LKJ correlation matrix \citep{lewandowski2009generating, stan2016stan}. However, these methods are not designed to handle the covariance of a single time series. For a single time series obeying eq. \eqref{eq:2}, there is just one multivariate data point (that is, the vector $\mathbf{y}$) available to inform the matrix $\Sigma$. With such limited data, standard methods for estimating covariance matrices have too much freedom, resulting in dense matrices that overfit the data.

Our strategy is to impose a positive definite covariance function $C : \mathbb{R} \times \mathbb{R} \rightarrow \mathbb{R}$, which generates a covariance matrix according to the rule,
\begin{equation}  \label{eq:kernel}
\Sigma_{ij} = C(t_i, t_j).
\end{equation}
For example, heteroscedastic errors, where $\Sigma=\text{diag}(\mathbf{f}(\theta) \sigma^2)$, could be represented by the following covariance function:
\begin{equation}
C(t_i, t_j) = f(t_i; \theta) \sigma^2 \delta_{ij}. 
\end{equation}
where $\delta_{ij}=1,\;\text{if } i=j; \; 0, \text{otherwise}$. In this paper, we consider positive definite kernels which are flexible enough to capture a wide variety of noise forms, with parameters that can, nonetheless, be learned from a single time series.

\subsection{Kernels for time series noise}
In this section, we introduce the kernels used throughout this paper. Notwithstanding the important differences discussed in \S\ref{gp_comparison}, much of the work on kernel functions for Gaussian processes is applicable to ODE noise models as well, and the three kernels we discuss have seen extensive use in Gaussian process regression. One of the most widely used positive definite kernels is the Gaussian kernel (also called the Radial Basis Function, or RBF)~\citep{fasshauer2011positive},
\begin{equation}  \label{eq:rbf}
C(t_i, t_j) = \sigma^2 e^{-(t_i-t_j)^2/2L^2}.
\end{equation}
We also consider the Laplacian kernel~\citep{feragen2015geodesic} for specifying time series autocovariances, since it more faithfully reproduces the types of persistence emergent from basic univariate time series models,
\begin{equation}  \label{eq:abel}
C(t_i, t_j) = \sigma^2 e^{- |t_i - t_j|/L}.
\end{equation}
The kernels in eqs. \eqref{eq:rbf}~\&~\eqref{eq:abel} are each characterised by two parameters which control the size and autocovariance in the errors. A more general class of kernels is the Mat\'ern~\citep{williams2006gaussian},
\begin{equation}
C(t_i, t_j) = \sigma^2 \frac{2^{1-\nu}}{\Gamma(\nu)} \left( \frac{\sqrt{2\nu}}{L} |t_i - t_j| \right)^\nu K_\nu \left( \frac{\sqrt{2\nu}}{L} |t_i - t_j| \right),
\end{equation}
where $K_\nu$ is the modified Bessel function of the second kind. For $\nu = 1/2$, the Mat\'ern kernel simplifies to the Laplacian kernel.

\subsection{Comparison to Gaussian processes (GPs)}  \label{gp_comparison}
Consider a function $g:\mathcal{X} \rightarrow \mathbb{R}$ obeying a GP with mean function $m$ and kernel $C$, i.e. $g \sim \mathcal{GP}(m, C)$ (see, for example, \cite{rasmussen2003gaussian}). For every finite set of inputs $\{t_i\}_{i=1}^N,\, t_i \in \mathcal{X}$, the vector of function values $\mathbf{g} = (g(t_1), \dots, g(t_N))^\top$ has a multivariate Gaussian distribution,
\begin{equation}  \label{eq:gpr}
\mathbf{g} \sim \mathcal{N}(\mathbf{m}, \Sigma),
\end{equation}
where $\mathbf{m}=(m(t_1),\dots,m(t_N))^\top$ and $\Sigma$ is generated as in eq. \eqref{eq:kernel}. This distribution, identical with eq. \eqref{eq:2} for $m(\cdot)=f(\,\cdot\,;\theta)$, illustrates an apparent resemblance between the multivariate normal likelihood for time series noise and the GP. Our proposed noise model, however, differs from a GP in several key aspects:
\begin{enumerate}
\item In GP regression, eq.~\eqref{eq:gpr} determines a \textit{prior} over functions, and the posterior over functions is inferred. Our proposed noise model uses the multivariate normal specification as a \textit{likelihood} for finite observed data, and posterior inference applies only to the parameters of $f$, not the functional form of the noise-free relationship between $y$ and $t$ which is assumed fixed and fully determined by $\theta$.
\item To handle noisy data, Gaussian process regression typically adds an extra noise term---often IID Gaussian. No such terms are used in our multivariate normal noise process.
\end{enumerate}

That is, in full, the likelihood for our multivariate normal model is given by eq.~\eqref{eq:2}, with covariance matrix given by eq.~\eqref{eq:kernel}. An example of the utility of the multivariate normal noise process is shown in Figure~\ref{fig:posterior_stat_kernel}. In this example, we show that the Laplacian kernel can faithfully capture autoregressive order 1 (AR(1)) noise in an ODE time series model, enabling accurate posterior inference for the ODE model parameters.

\section{Flexible noise for ODEs using Gaussian processes}  \label{gp}
In this section, we describe the first of two flexible noise processes which can learn effective covariance matrices from a time series. Standard positive definite kernels such as~eq. \eqref{eq:rbf} and eq. \eqref{eq:abel} are appropriate for simple covariance matrices. They are, however, stationary: depending only on the difference between two time points and not on absolute time. In this section, we consider models that allow kernel parameters (for example, $\sigma$ and $L$ in~eq.~\eqref{eq:abel}) to vary smoothly over time, allowing distinct sections of a time series to have different noise magnitudes and persistences. First, a brief overview of existing work on non-stationary covariance functions is provided in \S\ref{review}. In \S\ref{nonstat_laplace}, the non-stationary version of the Laplacian kernel is presented. In \S\ref{gp_inference}, inference for non-stationary kernel parameters is introduced, and in \S\ref{gp_parameter}, GP hyperparameter selection is discussed. In \S\ref{gp_results}, results are presented on synthetic data using the non-stationary Laplacian kernel.
\subsection{Background on non-stationary covariance functions} \label{review}
Non-stationary covariance functions have been used for spatial modelling and Gaussian process regression. Unlike stationary kernels which depend only on the distance between the two inputs, in the non-stationary case, the kernel shape itself must depend on the input location. This is expressed using the notation $k_s(u)$ for a kernel centred at location $s$ and evaluated at location $u$. For example, for the Laplacian kernel with one dimensional input $t$, we would take 
\begin{equation} \label{eq:abel_input_dependent}
k_t(u) = \sigma(t)^2 e^{- |t-u|/ L(t)},
\end{equation}
with the kernel parameters $\sigma$ and $L$ being functions of the kernel centre location $t$. 

If eq.~\eqref{eq:abel_input_dependent} is used to construct a covariance matrix, there are no guarantees that it will be positive definite. Instead, non-stationary modelling has relied on the following general formula for a non-stationary positive definite covariance function:
\begin{equation}  \label{eq:nonstat}
C(x_i, x_j) = \int_{\mathbb{R}^N} k_{x_i}(u) k_{x_j}(u) du,
\end{equation}
for inputs $x_i, x_j, u \in \mathbb{R}^N$~\citep{higdon1999non, paciorek2003nonstationary}. The non-stationary version of the Gaussian RBF covariance function can be derived from this formula, which has been used in non-stationary Gaussian process regression~\citep{gibbs1998bayesian, paciorek2004nonstationary}. To learn time-varying kernel parameters, a Gaussian process prior can be placed on each~\citep{paciorek2004nonstationary, heinonen2016non}.

\subsection{Non-stationary Laplacian covariance function}  \label{nonstat_laplace}
In this section, we present a non-stationary version of the Laplacian kernel. The techniques presented here are equally applicable to any appropriate positive definite kernel, however.

The one-dimensional non-stationary Laplacian covariance function is:
\begin{equation}  \label{eq:nonstat_abel}
C(t_i, t_j) = \sigma(t_i) \sigma(t_j) \sqrt{\frac{2 L(t_i)L(t_j)}{L(t_i)^2 + L(t_j)^2}} \exp\left(-\frac{|t_i-t_j|}{\sqrt{L(t_i)^2 + L(t_j)^2}}\right).
\end{equation}
Eq.~\eqref{eq:nonstat_abel} may be derived as a special case of the non-stationary Mat\'ern kernel~\citep{paciorek2004nonstationary}; it also follows directly from the one-dimensional case of eq. \eqref{eq:nonstat} using reparametrised versions of the respective stationary kernels, with the reparametrisations chosen to ensure that the final non-stationary covariance functions have a sensible form (cf. eq. (3.69) in \cite{gibbs1998bayesian}). The logarithms of $L$ and $\sigma$ each vary over time governed by Gaussian process priors:

\begin{equation}  \label{eq:nonstat_sqexp_prior}
\log L \sim \mathcal{GP}(\mu_l, K_l), \;\log \sigma  \sim \mathcal{GP}(\mu_\sigma, K_\sigma),
\end{equation}
where $\mu_l$, $K_l$, $\mu_\sigma$, and $K_\sigma$ are the GP hyperparameters.

\subsection{Inference}  \label{gp_inference}
Having specified a non-stationary covariance function such as eq.~\eqref{eq:nonstat_abel}, the next task is to infer the posterior distribution of model and covariance parameters. However, analytic expressions for the posterior mean and variance of the Gaussian processes $L(t)$ and $\sigma(t)$ are not available. Instead, MCMC sampling or maximum a posteriori (MAP) estimation can be used to infer the values of $L(t)$ and $\sigma(t)$ at each time point~\citep{heinonen2016non}. MCMC sampling yields a set of samples distributed according to the posterior distribution, while MAP estimation uses optimisation to find the parameters values at which the posterior distribution is maximised. For both MCMC and MAP estimation, we recommend the use of gradient-based methods (e.g., Hamiltonian MCMC and gradient-based optimisers) for improved convergence rates in the high dimensional parameter space~\citep{neal2011mcmc}. When analytic gradients are not available, automatic differentiation can be used. Indeed, all our GP examples presented here involve an interpolation scheme discussed in \S\ref{performance}, rendering analytic derivatives intractable, and we resort to using automatic differentiation. 

MCMC sampling is advantageous because it can, in theory, fully characterise posterior uncertainty in the covariance parameter fits. However, it is computationally costly and may be not be necessary for concentrated unimodel posteriors. In these cases, MAP estimation may be preferred due to its lower computational burden. While MAP estimates are often sufficient for the kernel parameter fits, we typically still require information about the posterior uncertainty in the ODE model parameters. Thus, we recommend the following procedure for long ODE time series problems with non-stationary covariance functions, which is specified in Algorithm \ref{alg:mcmc_map}. First, the joint MAP estimate of model parameters and covariance parameters is obtained using a gradient-based optimiser. Then, MCMC sampling is used to obtain the posterior distribution of model parameters conditional on the previously obtained MAP estimate of covariance parameters. For both optimisation and MCMC sampling, random or uniform initialisation of the covariance parameters will work for easier problems but will delay convergence on longer time series. In long time series problems with intelligible noise patterns, we recommend a data-driven initialisation of the covariance parameters in order to accelerate convergence of MCMC or optimisation. To initialise $L$ and $\sigma$ in the non-stationary Laplacian covariance function, we use the procedure given by Algorithm \ref{alg:mcmc_initialisation}. In practice, gradient-based optimisers such as L-BFGS-B~\citep{zhu1997algorithm} may settle at local maxima. Thus, we perform optimisation with multiple restarts, with each restart taking a different initial value. A set of variable yet plausible initial values for the restarts can be generated by rerunning Algorithm \ref{alg:mcmc_initialisation} multiple times with different sliding window widths.

\begin{algorithm}[H]
\SetAlgoLined

\KwIn{A parametric model $f(t; \theta)$, a non-stationary covariance function $C_\phi(t_i, t_j)$, and observed data $\{y_i\}_{i=1}^N$ at time points $\{t_i\}_{i=1}^N$; $\phi$ indicates the full set of kernel parameters defining $C$.}
\KwOut{MCMC samples distributed according to the conditional posterior $p(\theta | y, \phi_\text{MAP})$.}
  Initialise $\phi$, for example using to Algorithm \ref{alg:mcmc_initialisation} for the Laplacian kernel\;
  Use gradient-based optimisation to find $(\theta_\text{MAP}, \phi_\text{MAP}) = \text{arg max } p(\theta, \phi|y)$\;
  Calculate the fixed covariance matrix $\Sigma_\text{MAP}$ such that $\Sigma_{i,j} = C_{\phi_\text{MAP}}(t_i, t_j)$\;
  Use the covariance matrix defined above to form the likelihood $\mathcal{N}(y|f(t;\theta), \Sigma_\text{MAP})$\;
  Use MCMC to sample from the conditional posterior $p(\theta | y, \phi_\text{MAP})$
 \caption{MCMC estimates of ODE parameters using MAP estimates for kernel parameters.}\label{alg:mcmc_map}
\end{algorithm}

\begin{algorithm}[H]
\SetAlgoLined
\KwIn{A parametric model $f(t; \theta)$ and observed data $\{y_i\}_{i=1}^N$ at time points $\{t_i\}_{i=1}^N$ with spacing $\Delta t$. A sliding window width for each kernel parameter.}
\KwOut{A rough estimate of the parameters $\{\sigma_i\}_{i=1}^N$ and $\{L_i\}_{i=1}^N$ of the non-stationary Laplacian kernel which can be used to initialise an optimisation algorithm or MCMC sampler.}
  Use optimisation to find the MAP estimate of model parameters assuming an IID noise model, $\theta_{\text{MAP}, \text{IID}}$\;
  Subtract $f(t; \theta_{\text{MAP}, \text{IID}})$ from the observed data to obtain an estimate of the noise terms $\epsilon_i$\;
  At each time point $t_i$, calculate the empirical variance $v_i$ and 1st order autocorrelation $\rho_i$ of the noise terms within a sliding window centred on that time point\;
  Smooth both estimates using a Wiener filter~\citep{wiener1950extrapolation}\;
  At each time point, $t_i$, set $\sigma_i = \sqrt{v_i}$ and $L_i = -\Delta t / \log(|\rho_i|)$
 \caption{Initialisation for non-stationary Laplacian kernel parameters.}\label{alg:mcmc_initialisation}
\end{algorithm}

\subsection{Gaussian process hyperparameters} \label{gp_parameter}
For the GPs defined by eqs.~\eqref{eq:nonstat_sqexp_prior}, we used squared exponential kernels with constant mean functions~\citep{heinonen2016non}. With this assumption, there are six Gaussian process hyperparameters for the model (for each of $L$ and $\sigma$, a mean $\mu$, noise level $\alpha$, and length scale $\beta$). Prior knowledge or a grid search can be used to set these values, although existing work suggests only $\beta$ has a substantial effect on posteriors in most cases~\citep{heinonen2016non}. We set $\mu=0$ and $\alpha=1$~\citep{heinonen2016non}, and propose the following method to set $\beta$, which controls how the Gaussian process can change over time. This behaviour is crucial to the adaptivity of the method. If the scale, $\beta$, is too short, the GP will overfit local fluctuations; too large and it will fail to account for real changes in the process over time.

To set the length scale, we used a heuristic based on the expected rate of change of the noise process. Given evenly spaced time points with spacing $\Delta t$ and a user-specified number of time points $N_c$, we set $\beta$ as the solution of
\begin{equation}  \label{eq:gp_hyper}
\zeta = e^{-(N_c\Delta t)^2 / (2\beta^2)},
\end{equation}
for some small value $\zeta=0.01$. This equation imposes that the prior covariance between two values of the Gaussian process $N_c$ time points apart is close to zero, thus summarising the prior belief that the noise structure can change over that time scale. Good choices for $N_c$ will generally be problem specific. For non-uniform spacing, $N_c \Delta t$ could replaced by an appropriate time interval.

\subsection{Example with synthetic data}  \label{gp_results}
\begin{figure}[htbp]
\centering
\begin{subfigure}{\textwidth}
\centering
\includegraphics[width=.75\linewidth]{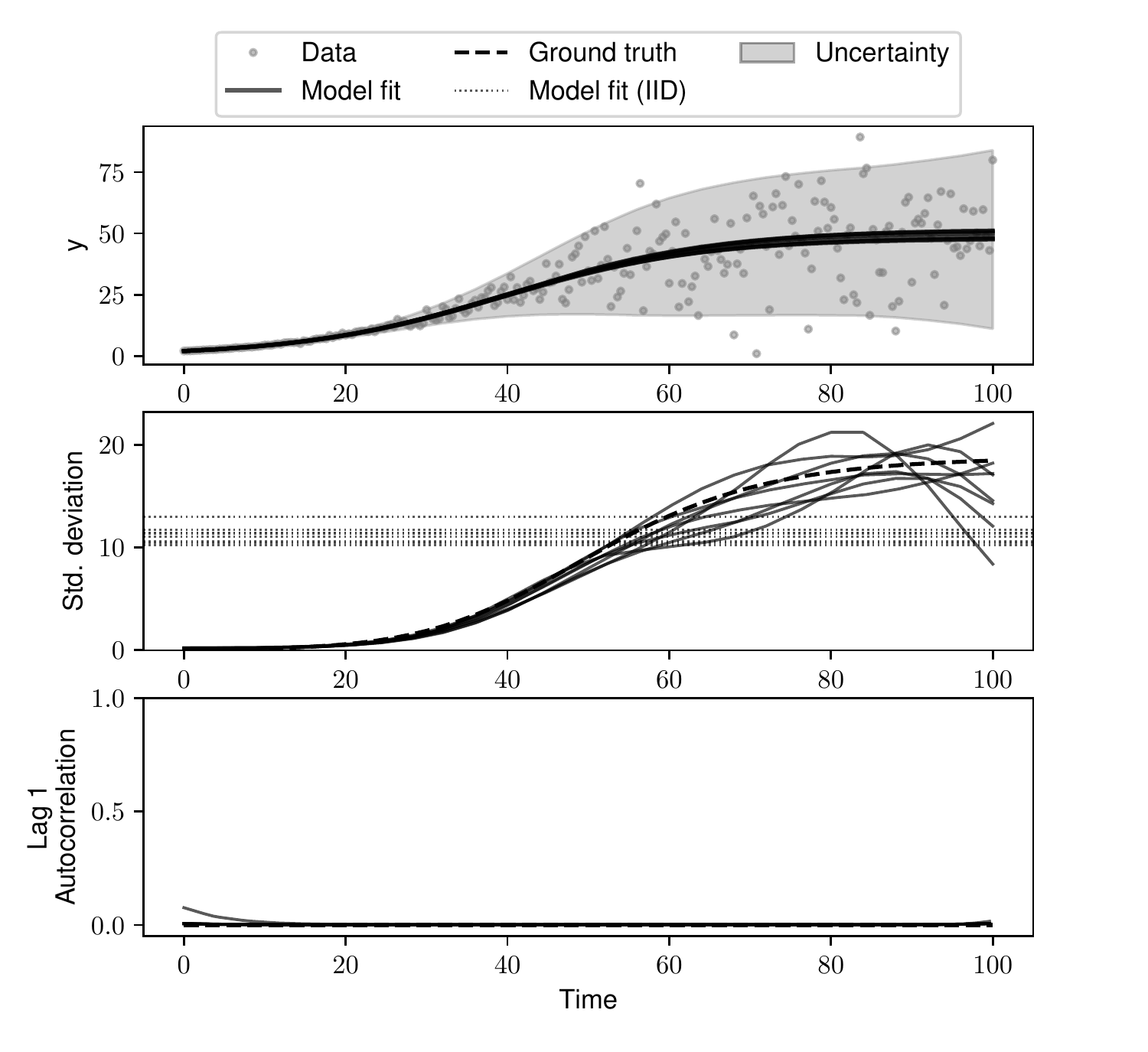}
\caption{}
\end{subfigure}
\begin{subfigure}{\textwidth}
\centering
\includegraphics[width=.75\linewidth]{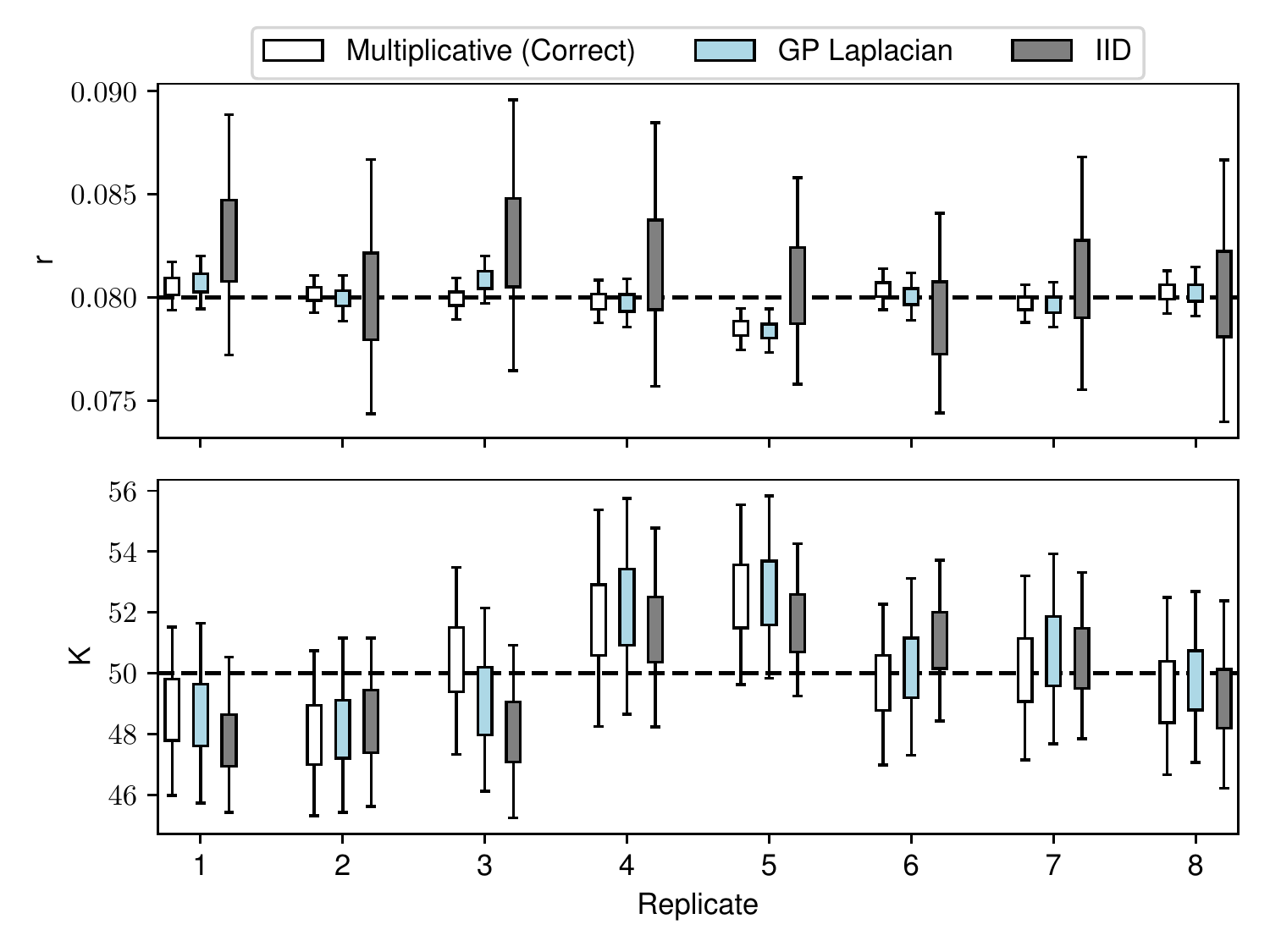}
\caption{}
\end{subfigure}
\caption{\textbf{Non-stationary Laplacian kernel fits to logistic data.} The top plot of panel (a) shows an example logistic growth time series with multiplicative noise, with 250 time points. In the other two plots of (a) and in panel (b), results for model fits to eight replicate datasets are shown. In the middle plot of panel (a), the true standard deviation $\sqrt{C(t_i, t_i)}$ is shown, along with model estimates of it at the MAP estimates for $L$ and $\sigma$ (one line per each replicate). In this plot, we also indicate the standard deviations estimated by the IID assumption as horizontal dashed lines. In the bottom plot of panel (a), the same is shown as in the middle plot, except with results for the lag 1 autocorrelation, $C(t_i, t_{i+1}) / (\sigma(t_i)\sigma(t_{i+1}))$. Panel (b) shows MCMC estimates of the posterior distributions for the logistic growth model parameters under three different assumptions for the noise process; the boxes cover the central 50\% posterior estimates, while the whiskers cover the central 95\% posterior estimates, and the dashed lines indicate the true values of the parameters.}
\label{fig:posterior_gp_kernel}
\end{figure}

In this example, we use the two-parameter logistic growth model:

\begin{equation}\label{eq:logistic}
\frac{dy}{dt} = r y (1 - y/K).
\end{equation}

We demonstrate the results of fitting the non-stationary kernel to synthetic data, generated from a logistic growth model with $r=0.08$, $K=50$, and $f(t=0)=2$ with multiplicative Gaussian noise:

\begin{equation}\label{eq:logistic_noise}
y_i = f(t_i; \theta) + f(t_i; \theta)^\eta v_i,
\end{equation}
where $y_i$ is an observed data point, $f(t;\theta)$ is the ODE model solution, and ${v_i\overset{\text{IID}}{\sim} \mathcal{N}(0, \sigma^2)}$. We set $\eta=2$ and $\sigma=0.0075$. Eqs. \eqref{eq:logistic}\&\eqref{eq:logistic_noise} were used to generate eight replicate time series, each with 250 time points. We considered parameter inference for each set of series under three different noise processes: multiplicative (i.e. the true noise process), the non-stationary Laplacian kernel, and IID Gaussian. In each case, Algorithm \ref{alg:mcmc_map} was used to generate posterior samples from $r$ and $K$. MCMC sampling for model parameters was performed using Pints inference software \citep{clerx2019probabilistic} with three Markov chains and a total of 20,000 iterations on each. The posterior was sufficiently simple here that a non-gradient-based sampler -- a form of adaptive covariance algorithm, called the Haario Bardenet method \citep{haario2001adaptive, johnstone2016uncertainty} -- was used opposed to Hamiltonian Monte Carlo. On a desktop processor, each chain took approximately 20 minutes to run. The first half of each chain was discarded as warm-up, and convergence was assessed using the Gelman $\hat{R}$ statistic, requiring $\hat{R}<1.05$ for all parameters~\citep{gelman2013bayesian}. To set the GP hyperparameter $\beta$, we used eq.~\eqref{eq:gp_hyper} with $N_c=200$. The results are shown in Figure~\ref{fig:posterior_gp_kernel}. In panel (a), the data (from the first replicate) is shown in the top panel, along with the fitted model trajectory. Below, the standard deviation and lag 1 autocorrelation are shown based on the MAP estimates for each replicate and indicate good correspondence with the ground truth. In panel (b), the posterior distributions for the model parameters are shown. The growth parameter, $r$, was most affected by incorrectly assuming IID Gaussian noise, where the IID noise model resulted in estimates with overly inflated uncertainty. This is because model output is most sensitive to $r$ in the first half of the series, where the IID noise model overestimates the noise level. In all cases, the GP method provided a higher fidelity estimate of uncertainty than IID noise; in most cases the location of the posterior is also improved. Another example of the GPs fitted to synthetic data is given in Figure~\ref{fig:gp_blocks}. In this example, the true data generating process consists of discrete blocks of different noise models, and the results show the ability of the non-stationary kernel method to find an appropriate smooth approximation.

\section{Flexible noise for ODEs using change points}  \label{blocks}
In this section, we describe a second flexible noise process for learning covariance matrices for time series noise. In \S\ref{block_intro}, an overview of our ``block covariance matrix'' method is provided along with a change point model. In \S\ref{block_results}, we use synthetic data from the logistic model to illustrate how the model can be fitted.

\subsection{Nonparametric change point models}  \label{block_intro}
Instead of assuming that kernel parameters vary smoothly and continuously over time, we now introduce a model that divides the time domain into discrete sections; each with a distinct noise model. A model of this type results in a block structure for the covariance matrix. Assuming the time series is divided into $M$ regimes indexed by $m=1,\dots,M$, and each regime $m$ has a covariance kernel $k_m$, the covariance matrix takes the form,
\begin{equation}  \label{eq:block_diag}
\Sigma = \begin{pmatrix} \Sigma^{(1)} & 0 & \dots & 0 \\ 0 & \Sigma^{(2)} & \dots & 0 \\ \vdots &  \vdots & \ddots  & \vdots \\0 & 0 & \dots & \Sigma^{(M)} \end{pmatrix},
\end{equation}
where each block $\Sigma^{(m)}$ is a positive definite matrix, with elements given by ${\Sigma^{(m)}_{ij} = k_m(t_{i}, t_{j})}$. The covariance matrix from eq. \eqref{eq:block_diag} is guaranteed to be positive definite because each block is positive definite. Eq.~\eqref{eq:block_diag} requires that the autocorrelation between consecutive points at the boundary between two blocks is zero. When autocorrelation varies more gradually, this may be a disadvantage, and could likely be better handled by the non-stationary kernel method in \S\ref{gp}. But, for more rapid changes in covariance, the block method should outperform.

To infer the locations of block boundaries is a type of ``change point'' problem (see, for example,~\cite{aminikhanghahi2017survey}), which is a wide field with many methods and fitting processes. We choose a nonparametric approach to change point detection, which has the benefit that the number of boundaries need not be fixed beforehand. In particular, we specify the ``restricted product partition model'' (PPM) induced by the Pitman-Yor process as a prior on the partitions~\citep{martinez2014nonparametric}. The PPM approach is convenient for learning model parameters, as the posterior over partitions can be inferred jointly with the posterior over ODE model parameters.

To specify the change point model, we place a prior over the partitions which lie in the class of set compositions of the index set $[N]=\{1, 2, \dots, N\}$, i.e., $\mathcal{C}_{[N]}$. This class contains all partitions $\{S_1, S_2, \dots, S_M\}$ such that $S_m = \{s_j + i : i =1, 2, \dots n_m\}$, where $s_0=0$, $s_j = n_1 +n_2 + \dots + n_j$, and $n_m$ is the number of indices in $S_m$ ~\citep{martinez2014nonparametric}. As an illustrative example, consider the case where $N=3$, in which $\mathcal{C}_{[3]}$ contains the following elements:
$$
\{\{1\}, \{2\}, \{3\} \}, \;\; \{\{1, 2\} , \{3\} \},\;\; \{\{1\}, \{2, 3\}\},\;\; \{\{1, 2, 3\}\}.
$$
$\mathcal{C}_{[N]}$ contains all admissible assignments of the time points to consecutive blocks. 

Let $\rho_N$ indicate a random variable taking values in $\mathcal{C}_{[N]}$. For a problem with $N$ time points and $k$ regimes, the PPM prior on $\rho_N$ with discount parameter $s$ and strength parameter $\phi$ is given by:

\begin{equation}  \label{eq:ppm}
p(\rho_N|s, \phi) = \left( \frac{N!}{k!}\frac{\prod_{i=1}^{k-1} (\phi + i s)}{(\phi+1)_{N-1\uparrow}} \prod_{j=1}^k \frac{(1-s)_{n_j-1\uparrow}}{n_j!} \right).
\end{equation}

In this equation, $x_{n\uparrow}$ denotes the rising factorial or Pochhammer function; i.e., \linebreak
$x_{n\uparrow} = (x)(x+1)\dots(x+n-1)$. Some visual analysis of the prior $p(\rho_n)$ is presented in Figure~\ref{fig:block_prior}; $\phi$ serves as a location parameter controlling the number of blocks, while $s$ is a concentration parameter controlling the variance in this number. 

We assume a parametric form for all block kernels $k_m$, such as the Laplacian kernel given in eq.~\eqref{eq:abel} (which takes two parameters). These parameters are constant within a block, but vary from block to block. The appropriate priors for the kernel parameters will depend on the form of the kernel. For the Laplacian kernel, we place diffuse normal priors on the logarithms of the parameters:
\begin{equation}
\log L \sim \mathcal{N}(-4, 4), \; \log \sigma \sim \mathcal{N}(0, 4).
\end{equation}

MCMC inference for the posterior is possible using Metropolis-Hastings steps in a split-merge-shuffle sampler, as used in Algorithm 2 of~\cite{martinez2014nonparametric}. Each iteration of this approach consists of a series of proposals that may be accepted or rejected:  first, there is a proposal that affects the number of blocks; this can be either a merge of two consecutive blocks or a split at some random point within an existing block; second, a proposal that keeps the number of blocks the same is used -- in this ``shuffle'' proposal, the boundary between two blocks is randomly moved somewhere within their limits. 

In prior work on inference for PPM models~\citep{martinez2014nonparametric}, the MCMC algorithm is restricted to the case where all within-block parameters can be integrated out, allowing calculation of the marginal posterior probability $p(\rho_N|y)\propto p(y|\rho_N) p(\rho_N)$ for any valid proposed partition $\rho_N$. In this work, we relax this assumption, so that we can only calculate the likelihood given explicit values for both the partition and the block parameters. This presents a challenge to inference, as evaluations of the marginal likelihood are essential to calculate the acceptance ratio of the merge and split proposals. In this work, we modify the split-merge-shuffle sampler to handle the non-conjugate case using the ideas of the Sequentially-Allocated Merge-Split (SAMS) Sampler for Dirichlet processes~\citep{dahl2005sequentially}. Briefly, when proposing a split, we simultaneously propose new values for the block parameters in the newly split block using a random walk. When proposing a merge, the current parameter values from the first of the two merged blocks are selected for use in the proposal. At each MCMC iteration, an additional Metropolis-Hastings step is performed to update the explicit block parameter values. The details of our MCMC algorithm are given in \S\ref{block_mcmc}.

A key benefit of a nonparametric approach for covariance matrix estimation is that it can learn an appropriate number of blocks from data. This flexibility has a cost, however -- in theory, each individual time point could be assigned to its own block. To avoid this outcome, informative priors on $s$ and $\phi$ can be specified. For $s$, we specify the prior: $s \sim\text{beta}(1,1)$. For $\phi$, we place a shifted gamma prior\footnote{If $X-c\sim \text{gamma}(a, b)$, then $X \sim \text{shifted-gamma}(a, b, c)$.}\enlargethispage{-\baselineskip} with parameters $(a, b, s)$. We choose the hyperparameters $a, b$ such that, at the prior mean of $s$, prior mass in the number of blocks is heavily concentrated at one block. Appropriate values for $a$ and $b$ may thus depend on the length of the time series and the desired strength of the prior preference for one block; we achieve a moderately informative prior using $a=0.01$ and $b=100$. This specification represents a weak null belief that the noise process is stationary, but allows the noise process flexibility to change if necessary.

\subsection{Example with synthetic data}  \label{block_results}
\begin{figure}[hbtp]
\centering
\begin{subfigure}{\textwidth}
\centering
\includegraphics[width=.7\linewidth]{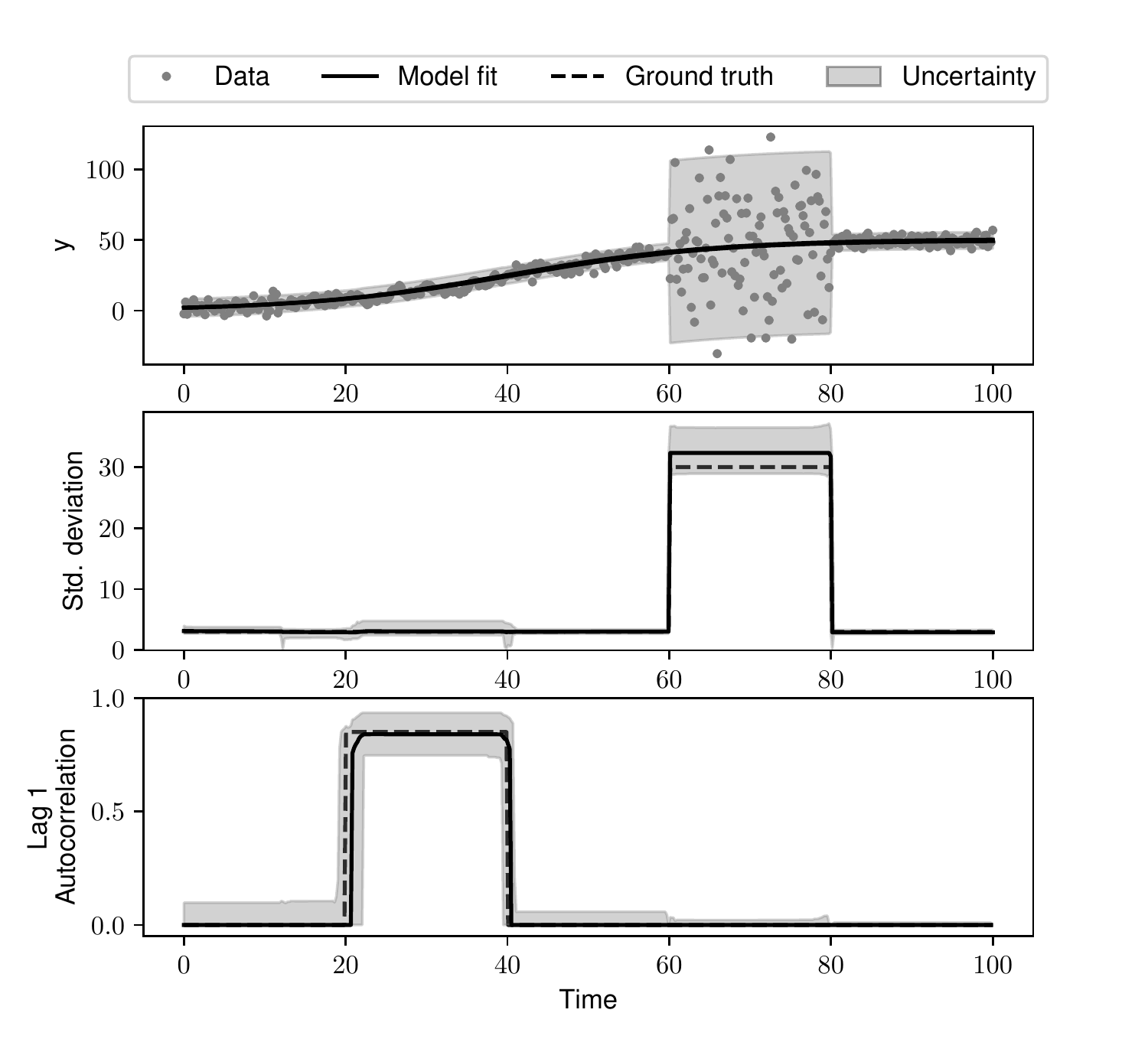}
\caption{}
\end{subfigure}
\begin{subfigure}{\textwidth}
\centering
\includegraphics[width=.7\linewidth]{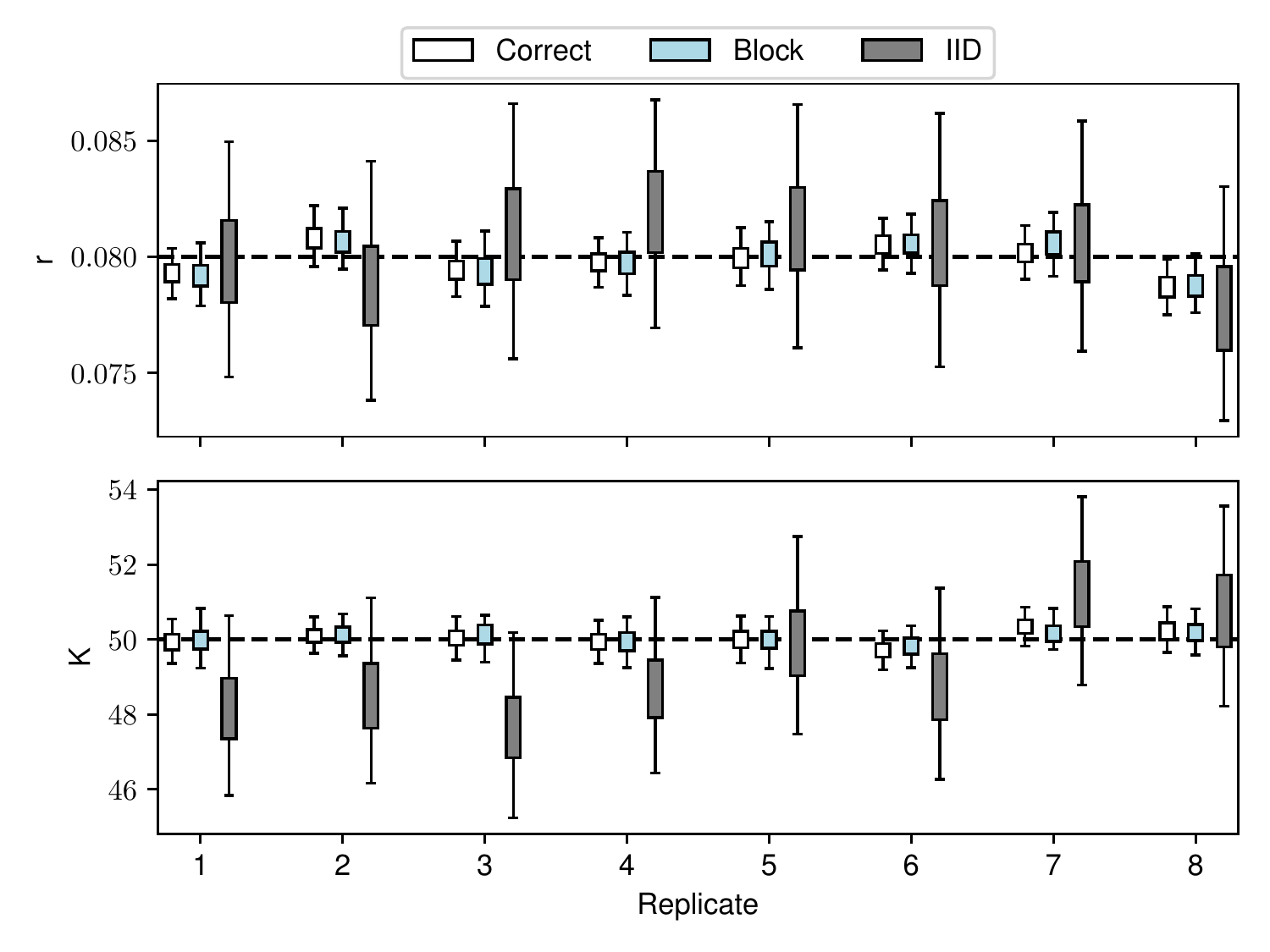}
\caption{}
\end{subfigure}
\caption{\textbf{Block covariance kernel fits to logistic data.} Panel (a) shows an example logistic growth time series with 5 blocks of different noise forms and a total of 500 time points. The model fit $\pm 2$ standard deviations of the inferred noise process are overlaid on the data points. In the middle and bottom plots of panel (a), the true values of standard deviation and lag 1 autocorrelation are shown as dotted lines, while the inferred posterior median standard deviation and central 90\% posterior are shown as a solid line and shaded region. In panel (b), results for model fits to eight replicate datasets are shown. This panel shows MCMC estimates of the posterior distributions for the logistic growth model parameters under three different assumptions for the noise process; the boxes cover the central 50\% posterior estimates, while the whiskers cover the central 95\% posterior estimates, and the dashed lines indicate the true values of the parameters.}
\label{fig:block_figure}
\end{figure}

We tested the block noise process on synthetic data generated according to the logistic model, eq.~\eqref{eq:logistic}, with $r=0.08$, $K=50$, and $f(t=0)=2$. In addition, we contrived a noise process with 5 regimes, each of 100 consecutive time points. The first, third and fifth regimes had IID Gaussian noise with $\sigma=3$, the second regime had AR(1) noise with $\rho=0.85$ and $\sigma=3$, and the fourth regime had IID Gaussian noise with $\sigma=30$. A total of 20,000 MCMC iterations were performed using our version of the split-merge-shuffle sampler (see \S\ref{block_mcmc}), with the first 10,000 discarded as warm up, and convergence was assessed using the Gelman $\hat{R}$ statistic, requiring $\hat{R}<1.05$ for all model parameters. Each MCMC iteration consisted of one split-merge-shuffle step for the blocks as well as one adaptive covariance step for the model parameters. On this dataset, all iterations took approximately 100 minutes to run on a desktop processor. In the first panel of Figure~\ref{fig:block_figure} (a), we show the data and the posterior median of the learned model trajectory, as well as the posterior median of two standard deviations of the inferred noise process about the model trajectory. The next two panels compare the estimated error standard deviation and lag 1 autocorrelation with their true values. This shows that the change point flexible noise process readily captures the five different regimes and learns their boundaries. In Figure~\ref{fig:block_figure} (b), we plot the posterior distributions for the logistic growth parameters with three different assumptions for the noise process. The noise process labelled ``Correct'' indicates an assumption of a multivariate normal likelihood with a block covariance matrix as given in eq.~\eqref{eq:block_diag}, with the block locations and sizes fixed to their correct values. However, in the ``Correct'' comparator method, the kernel parameters within each fixed block are not known, and are inferred jointly with the ODE model parameters. Meanwhile, the results from the nonparametric block covariance method are labelled with ``Block''. For both $r$ and $K$, the IID noise model results in inflated estimates of posterior variance, while both the block covariance method and the true noise model recover precise posteriors.

\section{Efficient computation for long time series}  \label{performance}
In real-life time series problems, we often encounter numbers of data points in the hundreds or thousands. In these cases, the computational cost of the methods mentioned above becomes a serious hindrance. In this section, we thus provide two computational strategies which can significantly decrease the runtime, and enable scaling to long time series.

The computational cost of the multivariate normal likelihood given in eq. \eqref{eq:2} is sensitive to the number of time series points, $n$: the covariance matrix, $\Sigma$, is $n \times n$, and the multivariate normal requires its inverse and determinant to be calculated. For highly sampled time series data, $n$ is large enough that these computations are impossible.

To scale to long time series, we take advantage of the relative sparsity of the covariance matrix. Any reasonable kernel, including the kernels we study in this paper, will generate matrices whose values are close to zero sufficiently far away from the diagonal. We truncate the entries in our covariance matrix, setting all those below a small threshold ($10^{-9}$) to zero. This results in a sparse matrix whose inverse and determinant can be computed using sparse Cholesky decomposition.

The non-stationary kernel for the GP method presents another scaling challenge as it requires inferring the value of the various GPs at each time point. For the non-stationary Laplacian kernel, this means that $L(t)$ and $\sigma(t)$ in eq. \eqref{eq:nonstat_sqexp_prior} are estimated for all $t$. For long time series, the number of parameters to infer then becomes prohibitive. To reduce this cost, we infer only the GP posterior on a sparser grid of time points. The GP functions are then interpolated to populate the covariance matrix at the original time points; here, we use linear interpolation but recognise that, if the GP value changes rapidly, more nuanced schemes may be appropriate. By introducing the interpolation step, the analytical calculation of the gradient of the multivariate normal likelihood becomes intractable. Therefore, in order to use gradient-based optimisers or MCMC samplers (such as L-BFGS-B and Hamiltonian Monte Carlo)~\citep{zhu1997algorithm, neal2011mcmc}, we rely on automatic differentiation.

The specific speedup enabled by these two computational approximations will vary greatly according to details of the problem at hand but, in our experience, can be quite dramatic. With a time series of length 150, we found that learning the non-stationary Laplacian kernel parameters at every fifth point and then interpolating resulted in a speedup of approximately 500\% at each MCMC iteration, and using sparse covariance matrices resulted in a speedup of approximately 4100\% for evaluation of the multivariate normal likelihood. On a typical desktop computer, these approximations enable reasonable runtimes for time series with lengths on the order of 10,000 points, as we demonstrate in the following section.

\section{Application to hERG channel kinetics}  \label{results}
The preceding examples of the flexible noise processes used synthetic data; in this section, we fit flexible noise processes to real data generated from experiments on the hERG potassium ion channel. This problem is challenging because the noise is clearly not IID, and, also because there may be misspecification of the underlying ODE model. In \S\ref{sec:herg_intro}, we provide a brief description of the hERG channel and a model used to investigate its behaviour and also describe experimental data generated for this system. In \S\ref{sec:herg_results}, we show how a flexible noise process can capture non-IID noise trends leading to different estimates of model parameters compared to those from an IID noise model.

The hERG channel time series are long (7700 time points, after $10 \times$ thinning), and we expect that the variation in the magnitude and autocorrelation of their noise terms can be captured using a continuously varying method. Thus, in this section we use the non-stationary covariance kernel method from \S\ref{gp} along with those modifications given in \S\ref{performance} to allow efficient computation.

\subsection{Description of hERG problem}\label{sec:herg_intro}
The \textit{human Ether-\`{a}-go-go-Related Gene} (hERG) encodes the alpha subunit of the potassium channel Kv11.1 that conducts the rapid delayed rectifier potassium current $I_\text{Kr}$. This current is of great importance in cardiac electrophysiology and safety pharmacology; reduction of $I_\text{Kr}$ by pharmaceutical compounds or mutations can induce fatal disturbances in cardiac rhythm. Interest in this model generally centres on understanding the current response of the hERG channel when a voltage stimuli $V$ is applied. The current can be described with a Hodgkin \& Huxley-style structure model~\citep{hodgkin1952quantitative} given by:
\begin{equation}\label{eq:ikr}
    I_\text{Kr} = g_\text{Kr} \cdot a \cdot r \cdot (V - E_\text{K}),
\end{equation}
where $g_\text{Kr}$ is the maximal conductance, and $E_\text{K}$ is the reversal potential (Nernst potential) for potassium ions which can be calculated directly from potassium concentrations using the Nernst equation.

The kinetic terms of the model, $a$ and $r$, are governed by:

\begin{align}
    \frac{\mbox{d}a}{\mbox{d}t} &= \frac{a_{\infty} - a}{\tau_{a}}, &
    \frac{\mbox{d}r}{\mbox{d}t} &= \frac{r_{\infty} - r}{\tau_{r}}, \\
    a_{\infty} &= \frac{k_1}{k_1 + k_2}, &
    r_{\infty} &= \frac{k_4}{k_3 + k_4}, \\
    \tau_{a} &= \frac{1}{k_1 + k_2}, &
    \tau_{r} &= \frac{1}{k_3 + k_4},
    \intertext{where,}
    k_1 &= p_1 \exp(p_2 V),    & k_3 &= p_5 \exp(p_6 V), \\
    k_2 &= p_3 \exp(-p_4 V),   & k_4 &= p_7 \exp(-p_8 V).
\end{align}
The model has 9 parameters $\theta=(g_\mathrm{Kr}, p_1, p_2, \dots, p_8 )$ to be inferred, all of which are positive. These parameters are the maximal conductance $g_\mathrm{Kr}$ [pS] and kinetic parameters $p_1, p_2, p_3, \cdots, p_8$ [s$^{-1}$, V$^{-1}$, s$^{-1}$, $\cdots$, V$^{-1}$].

Experimental data of the current are taken from a freely available dataset (Lei et al., 2019a; 2019b), where the voltage stimuli $V$ were designed for parametrising the model.

The logarithm-transformation was applied to all model parameters $\theta$, such that the transformed parameters $\phi = \log(\theta)$ are unconstrained. To account for the impact of this non-linear transformation on the posterior, a Jacobian transformation was applied. Priors for $\phi$ were selected using existing literature results (Lei et al., 2019a; 2019b), and, for each element of $\phi$, a weakly informative prior Gaussian distribution was used (see Table~\ref{table:herg_prior} for the prior hyperparameters).

\subsection{Results}\label{sec:herg_results}
\begin{figure}[h!]
\centering
\includegraphics[width=.9\linewidth]{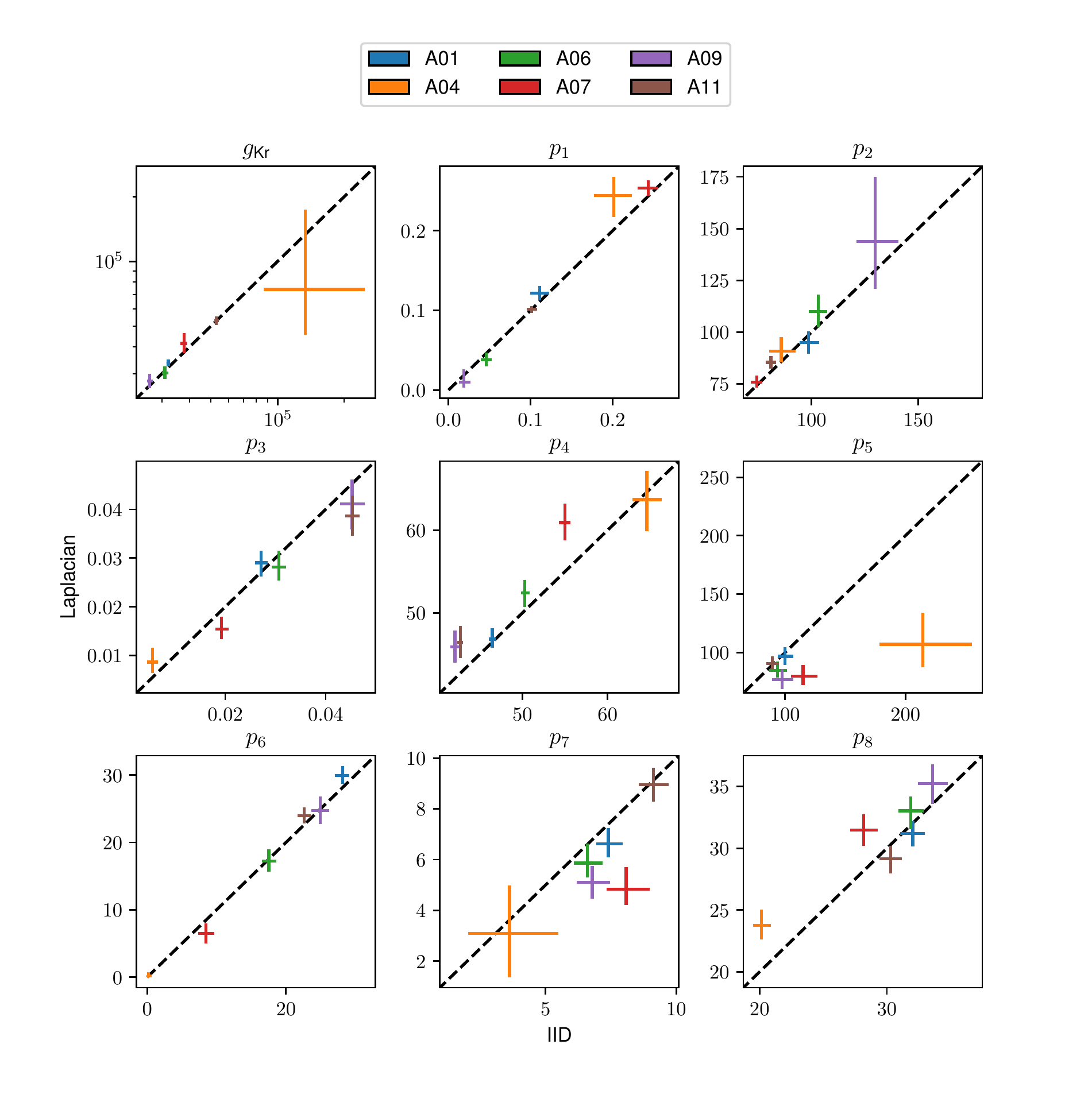}
\caption{\textbf{Posterior distributions for hERG model parameters}. This figure compares the posterior distributions resultant from the IID Gaussian noise assumption (``IID'') and non-stationary Laplacian kernel (``Laplacian'') for the nine hERG model parameters for six cells. For each parameter, the central 95\% range of the posterior is shown for each noise model as a bar, with the IID posterior shown on the horizontal axis and the non-stationary posterior shown on the vertical axis. Within each plot, a diagonal dashed line is drawn along $y=x$.}
\label{fig:block}   \label{fig:herg_posterior}
\end{figure}
\begin{figure}[h]
\centering
\begin{subfigure}{\textwidth}
\setcounter{subfigure}{3}
\centering
\includegraphics[width=.78\linewidth]{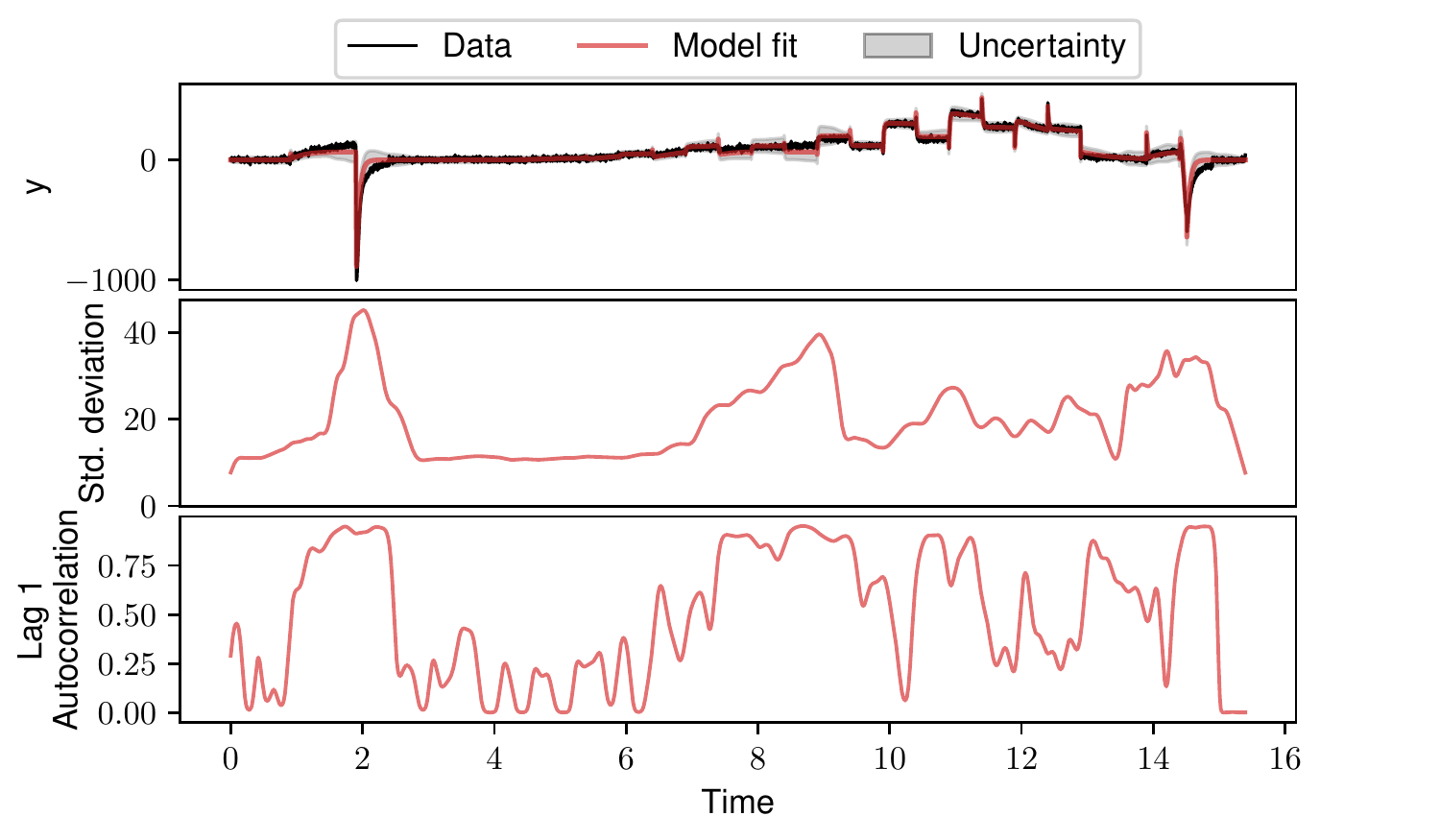}
\caption{A07}
\end{subfigure}
\begin{subfigure}{\textwidth}
\centering
\includegraphics[width=.78\linewidth]{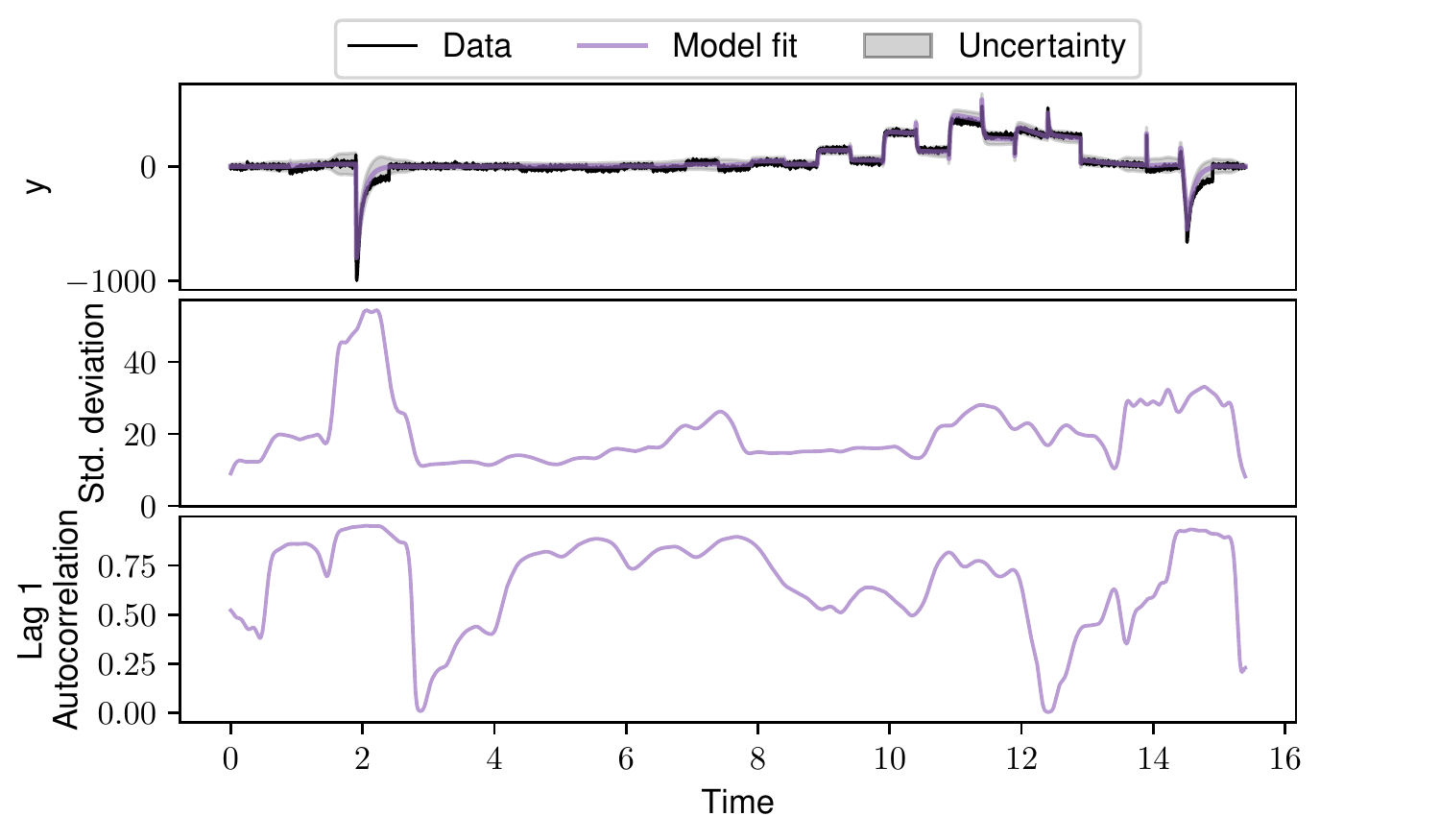}
\caption{A09}
\end{subfigure}
\begin{subfigure}{\textwidth}
\centering
\includegraphics[width=.78\linewidth]{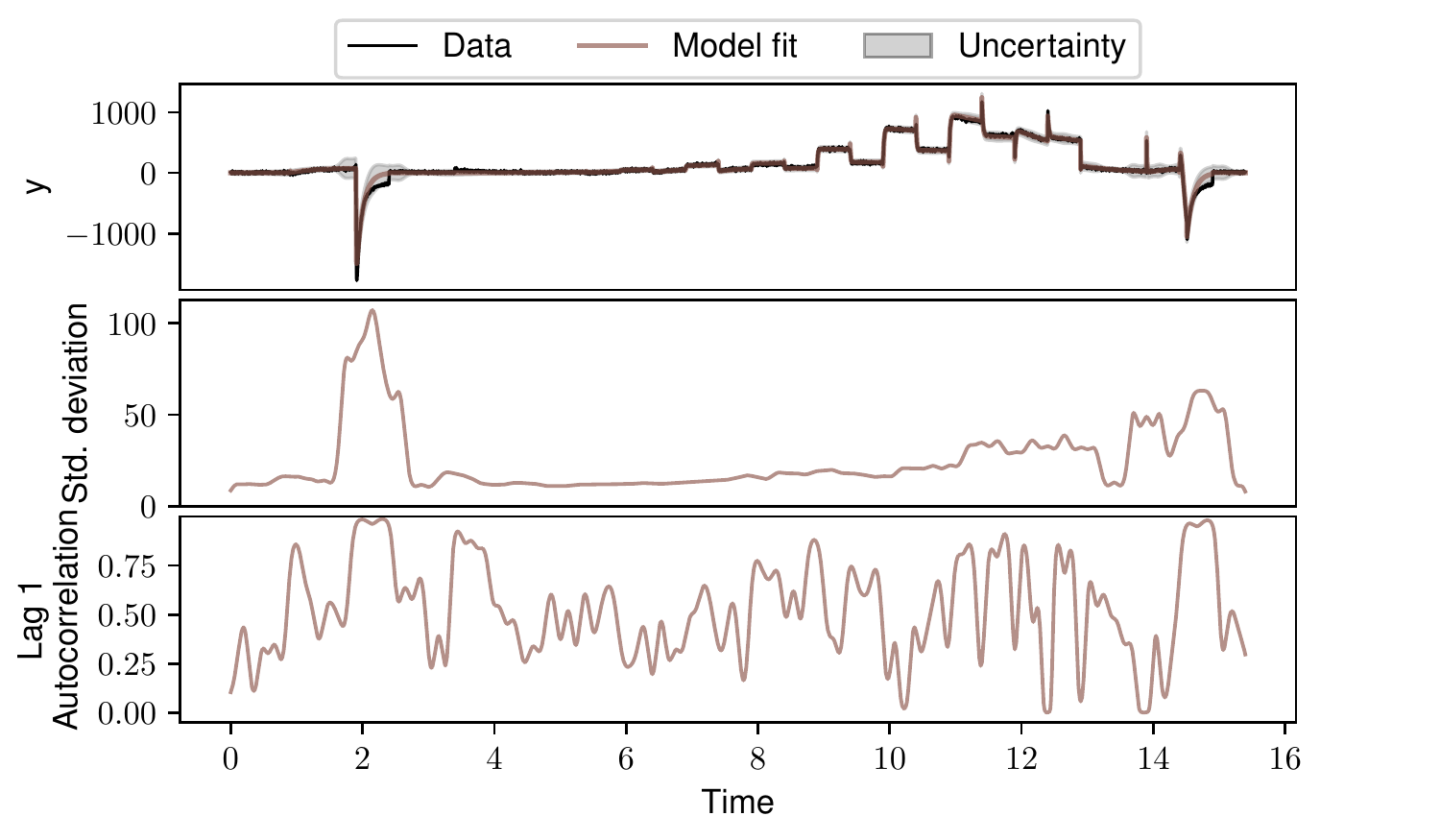}
\caption{A11}
\end{subfigure}
\caption{\textbf{Non-stationary Laplacian kernel noise models fit to hERG data}. This figure shows the data and model fit (top panel), and the MAP estimate standard deviation and lag 1 autocorrelation over time (second and third panels) inferred by the non-stationary Laplacian kernel noise model for each of six cells.}  \label{fig:herg}
\end{figure}

For six different cells, the model parameter posteriors were obtained via MCMC using the IID noise model and the non-stationary Laplacian kernel flexible noise model. To obtain posterior samples, the simulated tempering population MCMC algorithm was used~\citep{jasra2007population}, with convergence assessed using the Gelman $\hat{R}$ statistic~\citep{gelman2013bayesian}, and the first half of each chain discarded as warm up. For the non-stationary Laplacian kernel, we used Algorithm \ref{alg:mcmc_map} for inference and Algorithm \ref{alg:mcmc_initialisation} for initialisation. The fits for each of six cells are shown with the time series data in Figure~\ref{fig:herg}.

\begin{figure}[h!]
\centering
\includegraphics[width=.9\linewidth]{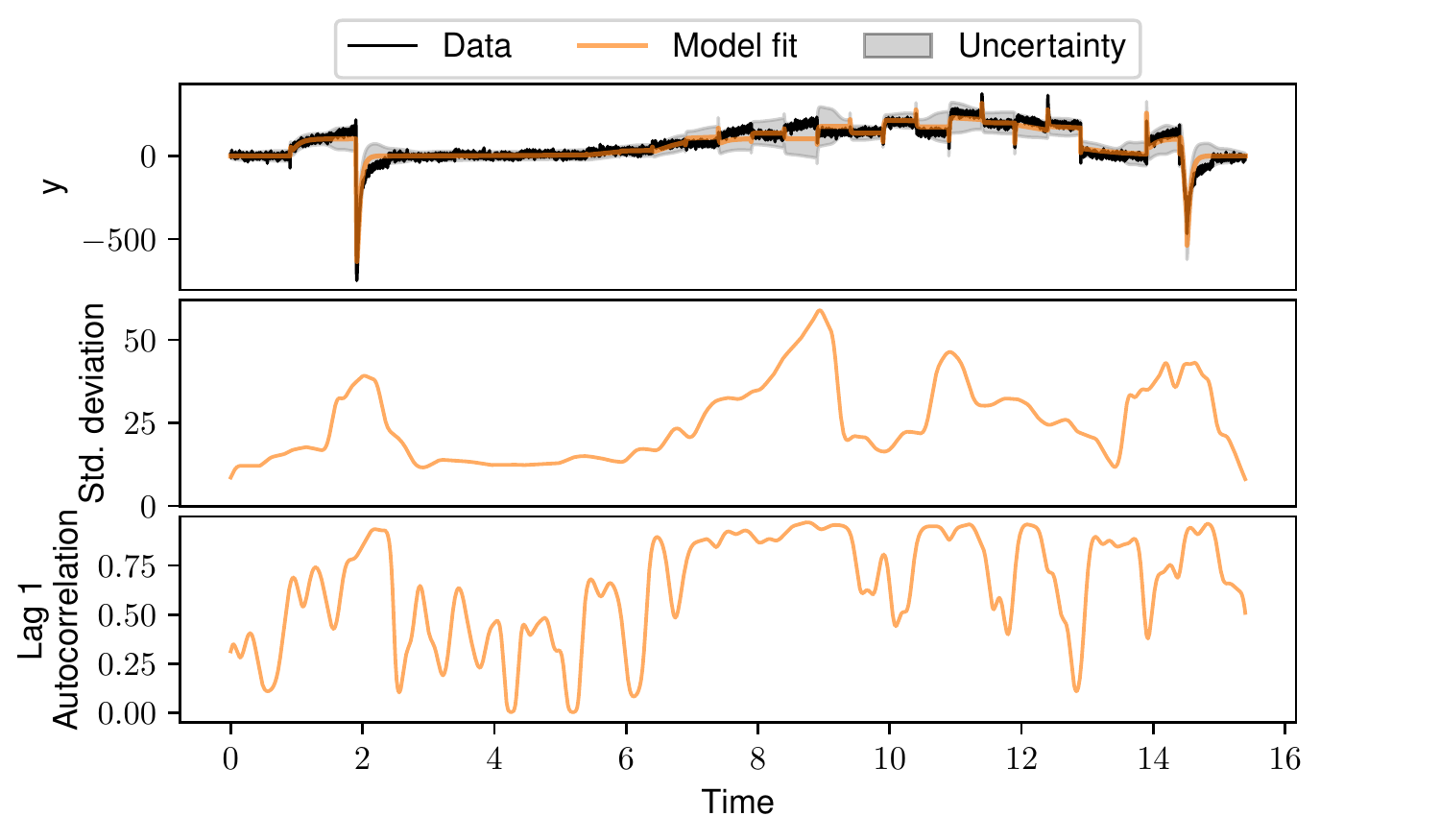}
\caption{\textbf{Non-stationary Laplacian kernel noise model fit to hERG cell A04.} This figure shows the data and model fit (top panel), and the MAP estimate standard deviation and lag 1 autocorrelation over time (second and third panels) inferred by the non-stationary Laplacian kernel noise model for cell A04.}
\label{fig:block}   \label{fig:herg_cell}
\end{figure}

Figure~\ref{fig:herg_posterior} shows the central 95\% posterior distribution ranges for all nine model parameters, assuming either IID Gaussian noise (horizontal axis) or the non-stationary noise process (vertical axis). There were significant differences in the parameter estimates for almost all parameters, with much of probability mass not overlapping the IID=Laplacian line. Additionally, the more sophisticated noise model resulted in substantially higher posterior variance for several model parameters, notably including $g_\text{Kr}$, $p_2$ and $p_4$. Cell A04 is an outlier: this is likely because this cell has a region of drastic misspecification in much of the time series, from $t=6$ to $t=10$. While the model fits for all six cells indicate short regions of misspecification, which is particularly apparent after the drops in current around $t=2$ and $t=14$, cell A04 (and to a lesser extent, A07) suffer from more extensive misspecification. The data and inferred fits for cell A04 are shown in Figure~\ref{fig:herg_cell}. The non-stationary noise model detects the central misspecified region by assigning high variance and autocorrelation in the middle of the time series. In the time series for cells A04 and A07, the poor fit between model and data may be largely explained by the fact that our model in this study (\S\ref{sec:herg_intro}) fails to account for experimental artefacts in the voltage clamp experiment, such as leakage current---these artefacts may explain much of the cell-to-cell variability observed in these experiments~\citep{lei2020accounting}. Thus, the high levels of standard deviation and autocorrelation detected in these time series suggest that a more detailed model of the experiment is necessary in order to understand these cells and correct the regions of obvious poor fit.

\section{Discussion}
When performing Bayesian inference for the parameters of time series models, the assumption made for the noise process may drastically alter the posterior estimates of parameter uncertainty. The flexible noise models described in this paper have the ability to learn noise processes from the data, including complex, non-stationary noise processes. The utility of these methods has been demonstrated in constructed synthetic data examples. 

In applied circumstances, noise terms which exhibit autocorrelation and time-varying magnitude often indicate model misspecification. This is what we observe in the hERG time series problem, in which the best fit model trajectory cannot fully express the signal that is clear in the data. In these cases, our non-stationary covariance noise process is able to pick out the regions of poor fit and model the spikes in magnitude and autocorrelation present at those time periods, with corresponding changes apparent in the model parameter posteriors. Future work to better handle misspecification in time series problems such as the hERG channel may benefit from the ability, offered by the methods in this paper, to avoid the often incorrect assumption of IID Gaussian noise.

\nocite{lei2019rapid}
\nocite{lei2019rapid2}

\section{Author contributions}
RC, BL, CLL, MR, and DG conceived the methods. RC designed the simulations and developed the algorithms and code. CLL set up the hERG data and model simulation code. RC, BL, and CLL wrote the manuscript. BL, MR, and DG supervised the analysis. All authors reviewed and approved the manuscript.

\section{Acknowledgements}
RC acknowledges support from the EPSRC. CLL acknowledges support from the Clarendon Scholarship Fund, and the EPSRC, MRC [EP/L016044/1] and F. Hoffmann-La Roche Ltd. for studentship support.

\bibliographystyle{chicago}
\bibliography{b}

\clearpage
\beginsupplement
\section{Stationary AR(1) noise with Laplacian kernel}
\begin{figure}[htbp]
\centering
\begin{subfigure}{\textwidth}
\centering
\includegraphics[width=.8\linewidth]{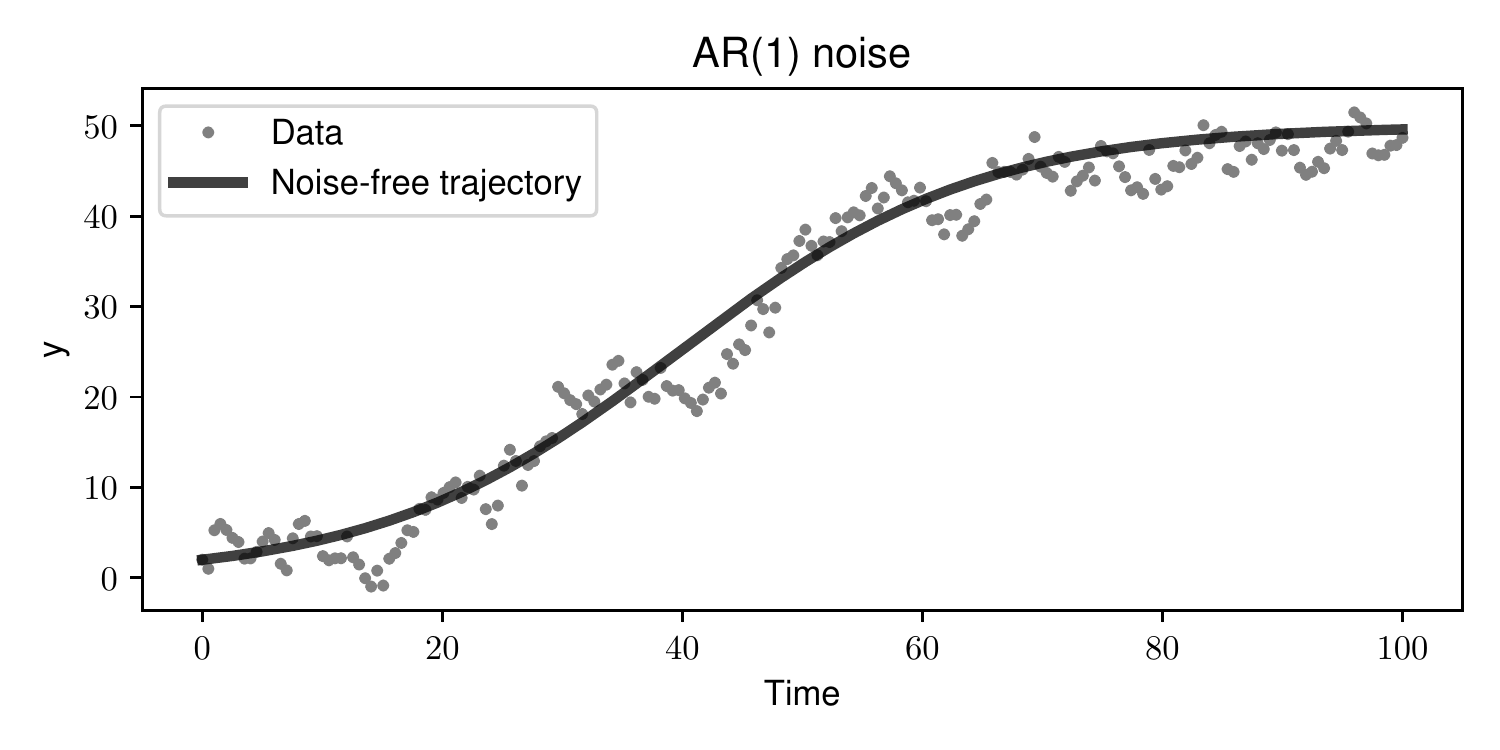}
\caption{}
\end{subfigure}
\begin{subfigure}{\textwidth}
\centering
\includegraphics[width=.8\linewidth]{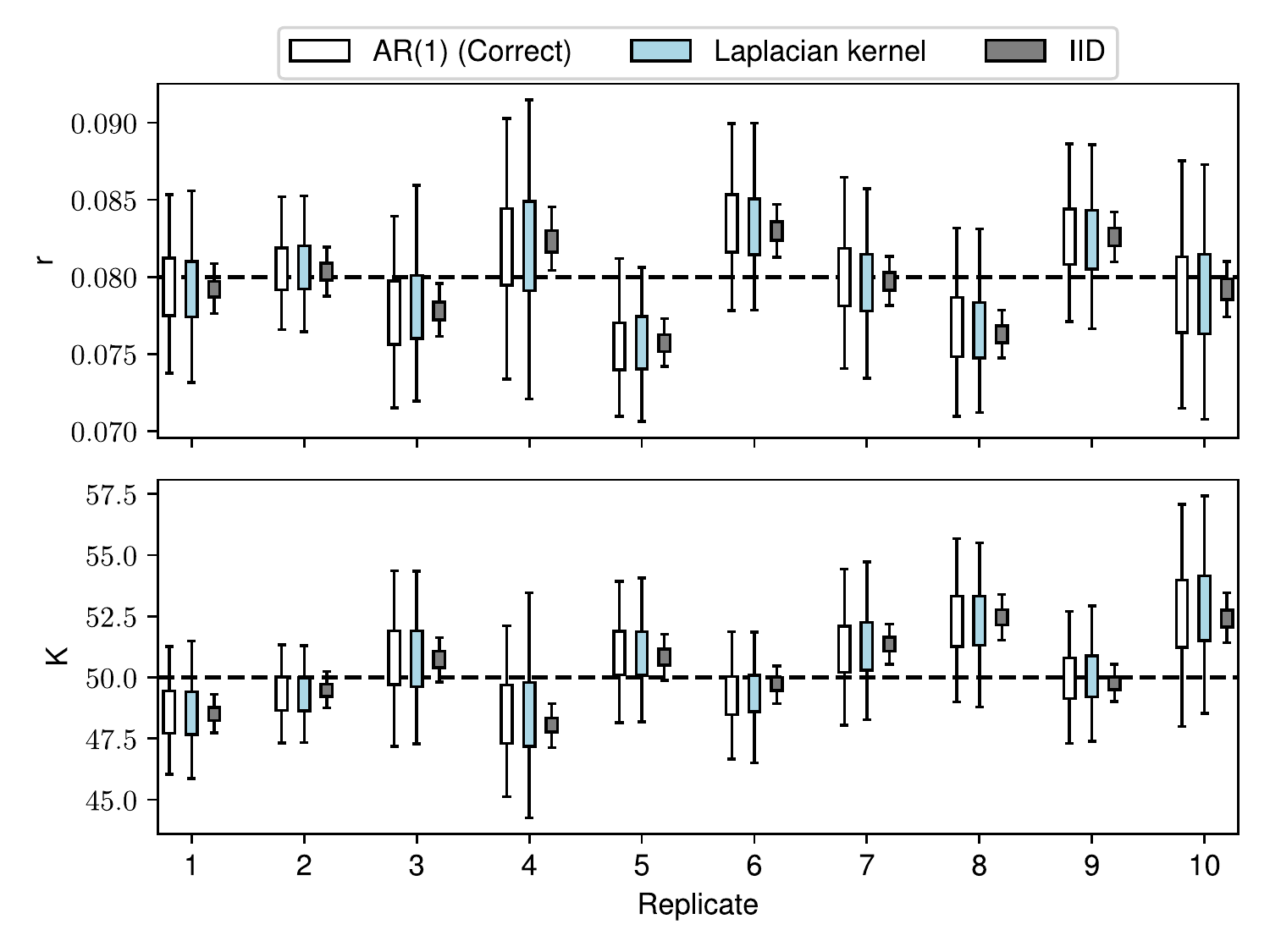}
\caption{}
\end{subfigure}
\caption{\textbf{Capturing AR(1) noise using a stationary Laplacian kernel}. Panel~(a) shows a logistic growth time series with AR(1) noise. Ten replicates of the AR(1) time series were generated. Panel~(b) shows the posterior distributions for logistic growth model parameters under three different assumptions for the noise process for each replicate. The boxes cover the central 50\% posterior estimates, while the whiskers cover the central 95\% posterior estimates. The dashed lines indicate true parameter values.}
\label{fig:posterior_stat_kernel}
\end{figure}
Accurate inference for stationary non-IID noise can be achieved using the standard Laplacian kernel,
\begin{equation} 
C(t_i, t_j) = \sigma^2 e^{- |t_i - t_j|/L}.
\end{equation}
Here, we show the results of the method applied to a synthetic logistic growth time series with autoregressive order 1 (AR(1)) error terms. These results are shown in Figure~\ref{fig:posterior_stat_kernel}. Panel (a) shows a synthetic noisy time series. The underlying model trajectory, labelled ``Noise-free trajectory'', is calculated from a logistic growth model, 
\begin{equation}
\frac{dy}{dt} = r y (1 - y/K).
\end{equation}
The AR(1) time series shows persistence in the error terms: the error term at any given time points depends both on a random fluctuation as well as the previous observation. Specifically, we model each error term $\epsilon_i = y_i - f(t_i;\theta)$ according to:
\begin{equation}
\epsilon_i = \rho \epsilon_{i-1} + v_i,
\end{equation}
where $v_i$ is Gaussian white noise, $v_i \sim N(0, \sigma \sqrt{1-\rho^2})$. In these simulations, we used $\rho=0.8$ and $\sigma=3$. Ten replicates of the time series with AR(1) noise were generated. For each time series, Bayesian inference for the parameters $r$ and $K$ was performed for each of three noise processes we consider: IID Gaussian with unknown variance (incorrectly specified), AR(1) with two unknown parameters (correctly specified), and the multivariate Gaussian likelihood with Laplacian kernel covariance. MCMC sampling was performed using three chains of the Haario Bardenet adaptive covariance algorithm, with a total of 20000 iterations in each chain~\citep{haario2001adaptive, johnstone2016uncertainty}. The first half of each chain was discarded as warm-up, and convergence was assessed using the Gelman $\hat{R}$ statistic~\citep{gelman2013bayesian}. In Panel (b), the results of posterior inference for $r$ and $K$ are shown under the three noise processes, across 10 replicates. In each replicate, the bars indicate the central 95\% of the posterior, while the green lines indicate the true values of the posterior. In each replicate, the first posterior with the correctly specified AR(1) noise process shows relatively high posterior uncertainty. The general multivariate normal noise process with Laplacian kernel reproduces the high level of posterior uncertainty in model parameters. By contrast, incorrectly specified IID assumption underestimates posterior uncertainty.

\section{Non-stationary Laplacian kernel on blocked synthetic data}
\begin{figure}[h!]
\centering
\includegraphics[width=.8\linewidth]{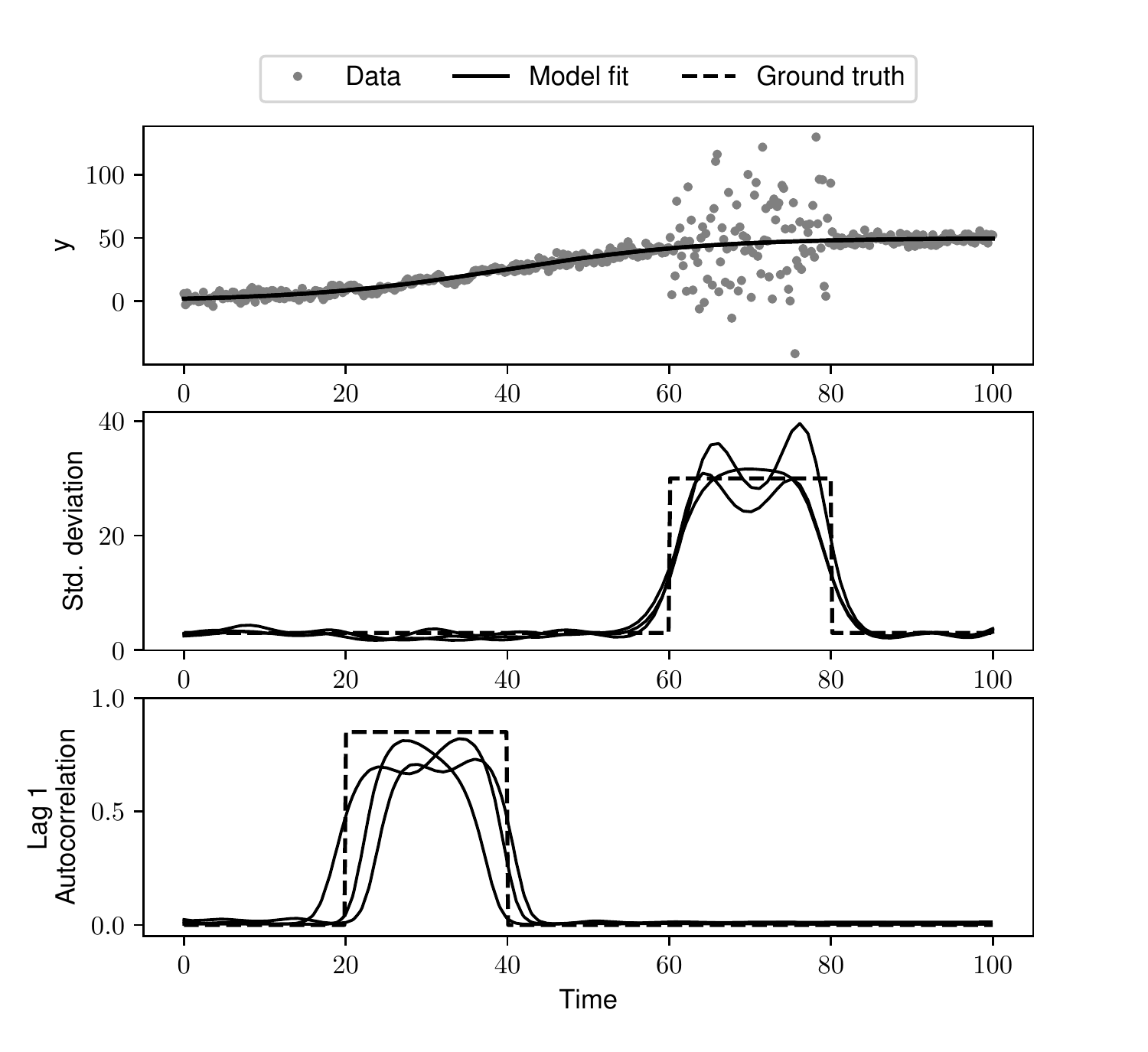}
\caption{\textbf{Non-stationary kernel method fit to blocked noise data}. This figure shows how the Gaussian processes in the non-stationary Laplacian kernel handle a noise process with blocks of different types of noise. The top plot shows a logistic growth time series with 5 blocks of different noise forms. In the middle and bottom plots, the true values of standard deviation and lag 1 autocorrelation are shown as dotted lines, while the inferred MAP estimates for standard deviation and lag 1 autocorrelation are shown in solid lines.}
\label{fig:gp_blocks}
\end{figure}
This section shows an example of the GP non-stationary kernel method being applied to a synthetic time series with very sharp changes in the true noise parameters. The noise process had 5 regimes, and was used with a logistic growth ODE model. The first, third and fifth regimes had IID Gaussian noise with $\sigma=3$, the second regime had AR(1) noise with $\rho=0.85$ and $\sigma=3$, and the fourth regime had IID Gaussian noise with $\sigma=30$. We found the MAP estimates of the non-stationary Laplacian kernel parameters, using Algorithm 2 for initialisation. In Figure~\ref{fig:gp_blocks}, the results are shown. The top panel shows one replicate of the data and the MAP estimate of the model trajectory. In the bottom two panels, the inferred standard deviation and lag 1 autocorrelation are shown for three replicates. The GPs are unable to learn the sharp corners in the ground truth for standard deviation and autocorrelation, but they do reach a smooth approximation of the ground truth.

\section{Blocked covariance prior distribution}
\begin{figure}[h!]
\centering
\includegraphics[width=.68\linewidth]{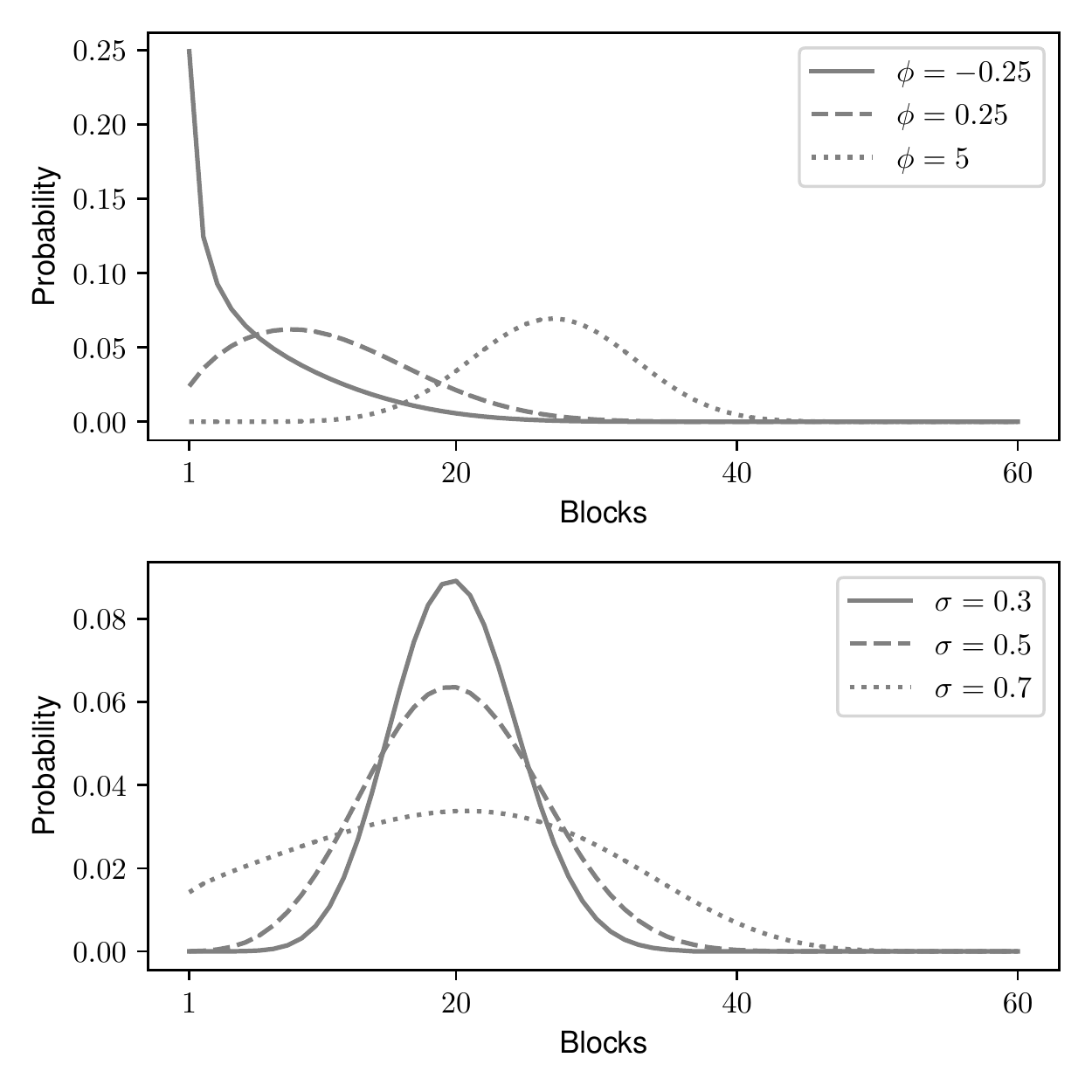}
\caption{\textbf{Marginal prior over number of blocks}. This figure shows slices of the marginal prior over the number of blocks used~\citep{martinez2014nonparametric}. In the top panel, $\sigma$ is fixed to $0.5$. In the bottom panel, the mean number of blocks is fixed to 20.}
\label{fig:block_prior}
\end{figure}
\noindent In this section, we present some visualisations of the PPM prior used in the block covariance method. The prior over partitions, given by
\begin{equation}  \label{eq:ppm1}
p(\rho_N) = \left( \frac{N!}{k!}\frac{\prod_{i=1}^{k-1} (\phi + i s)}{(\phi+1)_{N-1\uparrow}} \prod_{j=1}^k \frac{(1-s)_{n_j-1\uparrow}}{n_j!} \right),
\end{equation}
does not admit direct interpretation or visualisation, but the marginal prior distribution over the number of blocks offers insight into~\eqref{eq:ppm1} and its hyperparameters~\citep{martinez2014nonparametric}. Some slices of the marginal distribution over the number of blocks used are shown in Figure~\ref{fig:block_prior}. In all cases, the total number of time points $N$ is fixed to $60$. In the top panel, $s$ is fixed to $0.5$ and the prior is illustrated for three values of $\phi$. This shows how $\phi$ works as a location parameter, shifting the prior mass to higher numbers of blocks as it increases. In the bottom panel, the mean number of blocks is fixed to $20$ and the prior is shown for three values of $s$, which clearly serves as a shape parameter controlling the spread of the prior distribution.

\clearpage
\section{MCMC algorithm for block covariance} \label{block_mcmc}
This section describes our split-merge-shuffle MCMC algorithm for inferring blocked covariance matrices. The algorithm closely follows the split-merge-shuffle algorithm for the conjugate case~\citep{martinez2014nonparametric}, modified for non-conjugacy using ideas from the SAMS sampler~\citep{dahl2005sequentially}. We assume as input a parametric model $f(t; \theta)$, a positive definite kernel $C_{\psi_j}$, observed data $\{y_i\}_{i=1}^N$ at time points $\{t_i\}_{i=1}^N$, partition prior hyperparameters $s$ and $\phi$, with $\psi_j$ indicating the full set of kernel parameters defining $C$ in the $j$th block. Each data point is assigned to a block using indicator variables $\{z_i\}_{i=1}^N$. The steps for the sampler are provided in Algorithm~\ref{alg:mcmc_block}. 

To calculate $\alpha_\text{split}$, the acceptance probability for a proposed split, we must first propose a new value for $\psi$ in the new block. Following~\cite{dahl2005sequentially}, we propose $\psi_\text{split}$ according to a random walk, using a Gaussian proposal centred on the current value with a fixed variance. Letting $\rho$ and $\rho^*$ indicate the original and proposed split partitions, and $\psi$ and $\psi^*$ the original and proposed vectors of block kernel parameters, the acceptance ratio can then be calculated according to:
\begin{equation}  \label{eq:accept_ratio}
\alpha_\text{split} = \text{min}\left(1, \frac{p(\rho^*, \psi^* | y)}{p(\rho, \psi|y)} \frac{q(\rho, \psi|\rho^*, \psi^*)}{q(\rho^*, \psi^*|\rho, \psi)}  \right),
\end{equation}
where $p(\rho, \psi|y)$ denotes the posterior density evaluated at the partition $\rho$ and block kernel parameters $\psi$, while $q(\rho, \psi|\rho^* \psi^*)$ is the probability of proposing $\rho$ and $\psi$ from $\rho^*$ and $\psi^*$.
The basic forward and reverse proposal probabilities for the split are obtained in~\cite{martinez2014nonparametric}. Here, the forward probability must be multiplied by the proposal density used to obtain $\psi_\text{split}$. $\alpha_\text{merge}$ indicates the acceptance probability for a proposed merge. The same eq.~\eqref{eq:accept_ratio} applies, while in this case the reverse probability must be multiplied by the proposal density for $\psi_\text{split}$. The shuffle step is simpler as the dimension remains unchanged; the proposal probabilities for shuffling are taken from~\cite{martinez2014nonparametric}.

{\footnotesize
\begin{algorithm}[H]
\SetAlgoLined	
\KwIn{A parametric model $f(t; \theta)$, a positive definite kernel function $C_{\psi_j}$, observed data $\{y_i\}_{i=1}^N$ at time points $\{t_i\}_{i=1}^N$. Initial values for the following parameters: $\theta$, assignments $\{z_i\}_{i=1}^N$ of each time point to blocks, $\psi_j$ for each initial block $j$, and partition prior hyperparameters $s$ and $\phi$. Desired number of MCMC iterations $N_\text{MCMC}$.}

\KwOut{MCMC samples for $\theta$, $z$, $s$, $\phi$, and $\psi$.}

$K \leftarrow $ initial number of blocks\;
\For{$j=1\dots K$}  
{
$n_j \leftarrow $ $\sum_{i=1}^N  \mathbf{1}(z_i = j)$\tcp*{Count number of points in each block}
}
\For{$m=1\dots N_\text{\upshape MCMC}$}
{
Update $\theta$ via one adaptive covariance MCMC step\;
Update $s$ and $\phi$ via standard MH steps\;
\For{$j=1\dots K$}
{
Update $\psi_j$ via standard MH steps\;
}
Draw $u$ from $\text{uniform}(0, 1)$\;
\eIf{$K=1$ $\text{\upshape or}$ $u < 0.25$}
{
$z, \psi, K \leftarrow \texttt{split}(z, \psi, K, n)$\tcp*{Run the split routine (Alg.\ \ref{alg:split})}
}
{
$z, \psi, K \leftarrow \texttt{merge}(z, \psi, K, n)$\tcp*{Run the merge routine (Alg.\ \ref{alg:merge})}
}
\For{$j=1\dots K$}
{
$n_j \leftarrow $ $\sum_{i=1}^N  \mathbf{1}(z_i = j)$\tcp*{Recount number of points in each block}
}
$z \leftarrow \texttt{shuffle}(z, K, n)$\tcp*{Run the shuffle routine (Alg.\ \ref{alg:shuffle})}
}
\caption{MCMC sampler for block covariance}\label{alg:mcmc_block}
\end{algorithm}
}

\begin{algorithm}[H]
\KwIn{Assignments $z$, Kernel block parameters $\psi$, number of blocks $K$, number of points in each block $n$.}

\KwOut{Updated assignments $z$, block parameters $\psi$ and number of blocks $K$ after one Metropolis-Hastings split step.}

  \SetKwFunction{FMain}{split}
  \SetKwProg{Fn}{Function}{:}{}
  
   \Fn{\FMain{$z$, $\psi$, $K$, $n$}}{

Randomly select a block $j$ from $\{j : 1 \leq j \leq K , n_j > 1\}$\;
Randomly select an index $l$ from $\{1, \dots, n_j-1\}$\;
Draw $u'$ from $\text{uniform}(0, 1)$\;
\If{$u' < \alpha_\text{split}$}
{
\For{$i=1\dots N$}
{
\If{$i > l + \sum_{j'=1}^{j-1} n_{j'} $}
{
$z_i \leftarrow z_i + 1$\;
}
}
$\psi \leftarrow (\psi_1, \dots, \psi_j, \psi_\text{split}, \psi_{j+1}, \dots, \psi_K)$\;
$K \leftarrow K + 1$\;
}

\KwRet $z$, $\psi$, $K$\;

}
\caption{Split proposal step}  \label{alg:split}
\end{algorithm}

\begin{algorithm}[H]

\KwIn{Assignments $z$, kernel block parameters $\psi$, number of blocks $K$, number of points in each block $n$.}

\KwOut{Updated assignments $z$, block parameters $\psi$ and number of blocks $K$ after one Metropolis-Hastings merge step.}

  \SetKwFunction{FMain}{merge}
  \SetKwProg{Fn}{Function}{:}{}
  
   \Fn{\FMain{$z$, $\psi$, $K$, $n$}}{

Randomly select a block $j$ from $\{j : 1 \leq j \leq K-1\}$\;
Draw $u'$ from $\text{uniform}(0, 1)$\;
\If{$u' < \alpha_\text{merge}$}
{
\For{$i=1\dots N$}
{
\If{$i > \sum_{j'=1}^{j} n_{j'} $}
{
$z_i \leftarrow z_i - 1$\;
}
}
$\psi \leftarrow (\psi_1, \dots, \psi_j, \psi_{j+2}, \dots, \psi_K)$\;
$K \leftarrow K - 1$\;
}

\KwRet $z$, $\psi$, $K$\;

}
\caption{Merge proposal step}  \label{alg:merge}
\end{algorithm}

\begin{algorithm}[H]
\KwIn{Assignments $z$, number of blocks $K$, number of points in each block $n$.}
\KwOut{Updated assignments $z$ after one Metropolis-Hastings shuffle step.}

  \SetKwFunction{FMain}{shuffle}
  \SetKwProg{Fn}{Function}{:}{}
  
   \Fn{\FMain{$z$, $K$, $n$}}{

Randomly select a block $j$ from $\{j : 1 \leq j \leq K-1\}$\;
Randomly select an index $l$ from $\{1, \dots, n_j + n_{j+1} -1 \}$\;
Draw $u'$ from $\text{uniform}(0, 1)$\;
\If{$u' < \alpha_\text{shuffle}$}
{
\For{$i=1\dots N$}
{
\If{$\sum_{j'=1}^{j-1} n_{j'} < i \leq l + \sum_{j'=1}^{j-1} n_{j'} $}{
$z_i \leftarrow j$\;
}
\If{$l + \sum_{j'=1}^{j-1} n_{j'} < i \leq \sum_{j'=1}^{j+1} n_{j'}$}{
$z_i \leftarrow j+1$\;
}
}
}
\KwRet $z$\;
}
\caption{Shuffle proposal step}  \label{alg:shuffle}
\end{algorithm}

\section{hERG Hodgkin-Huxley model parameter priors}
In this section, we list the priors used for the 9 log-transformed model parameters in the hERG model introduced in \S\ref{sec:herg_intro}.

\begin{figure}[h]
\renewcommand\figurename{Table}
\setcounter{figure}{0}
\centering
\begin{tabular}{|c|c|} 
\hline
Parameter & Prior \\ \hline
$g_\mathrm{Kr}$ & $N(10.5, 1.0)$ \\ \hline
$p_1$ & $\mathcal{N}(-2.5, 3.0)$ \\ \hline
$p_2$ & $\mathcal{N}(4.5, 1.0)$ \\ \hline
$p_3$ & $\mathcal{N}(-3.5, 1.5)$ \\ \hline  
$p_4$ & $\mathcal{N}(4.0, 0.5)$ \\ \hline  
$p_5$ & $\mathcal{N}(4.5, 0.5)$ \\ \hline  
$p_6$ & $\mathcal{N}(3.0, 1.5)$ \\ \hline  
$p_7$ & $\mathcal{N}(2.0, 0.5)$ \\ \hline  
$p_8$ & $\mathcal{N}(3.5, 0.5)$ \\ \hline   \hline  
\end{tabular}
\caption{\textbf{hERG model prior parameters}. This table contains the prior distributions used for each parameter in the hERG model. For each parameter, the prior is a normal distribution with the mean and standard deviation given in the table.}  \label{table:herg_prior}
\end{figure}

\section{hERG data and model fits} \label{herg_results_supp}
\begin{figure}[h]
\setcounter{figure}{2}
\centering
\begin{subfigure}{\textwidth}
\centering
\includegraphics[width=.78\linewidth]{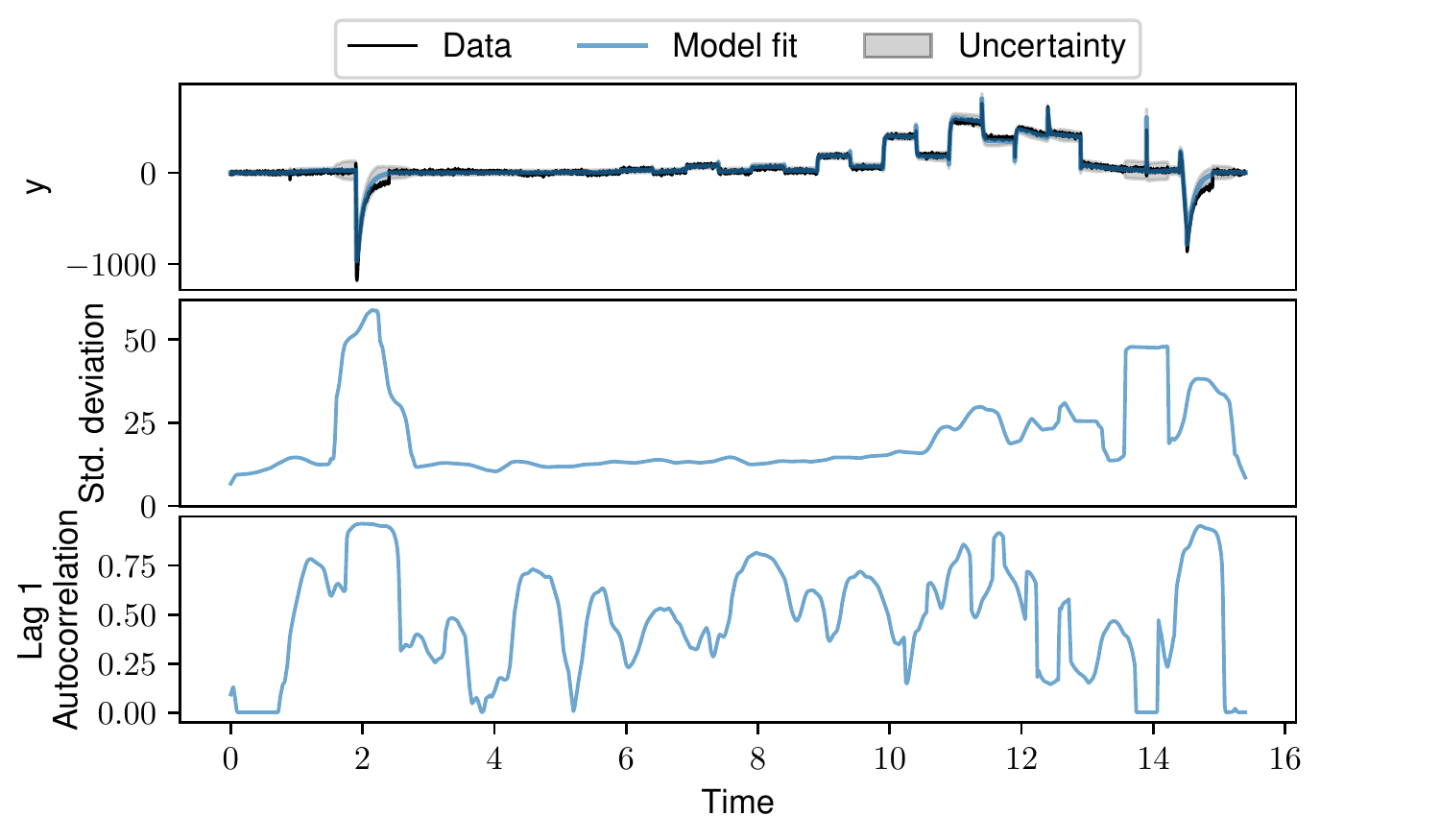}
\caption{A01}
\end{subfigure}
\begin{subfigure}{\textwidth}
\centering
\includegraphics[width=.78\linewidth]{A04-figs.pdf}
\caption{A04}
\end{subfigure}
\begin{subfigure}{\textwidth}
\centering
\includegraphics[width=.78\linewidth]{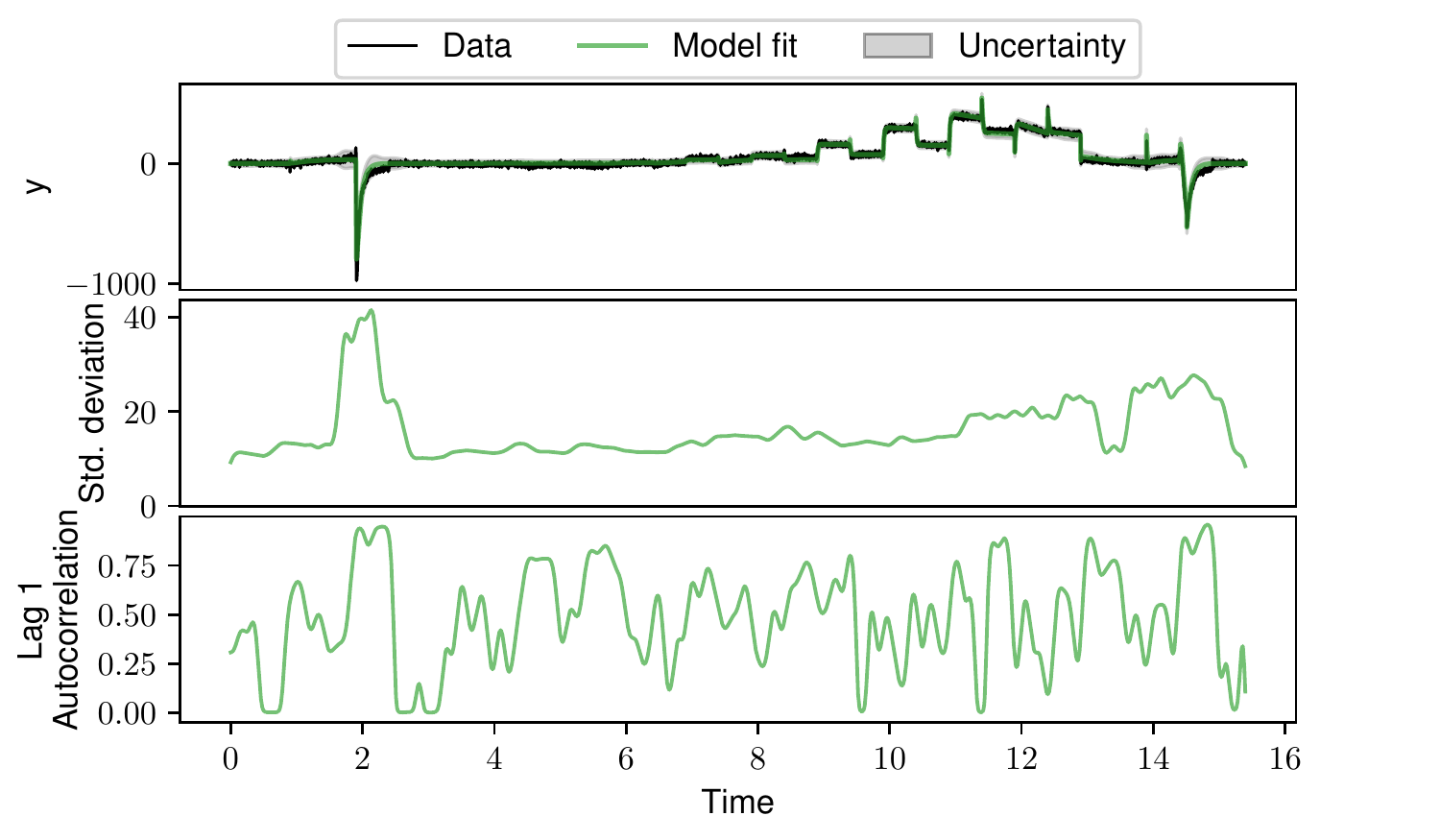}
\caption{A06}
\end{subfigure}
\end{figure}

\noindent This section contains the time series data and model fit for six hERG cells.  The main features of the Gaussian process fits are two large spikes in standard deviation appearing around times $t=2$ and $t=14.5$. While there is some variation from cell to cell, these spikes correspond to regions of the time series with rapid drops in current. Autocorrelation is fairly high throughout the time series, with wide fluctuations. Cells A04 and A07 exhibit another peak in standard deviation and autocorrelation around $t=8$; this corresponds to a region of obvious discrepancy between the model and the data.

\end{document}